\documentclass{SciPost}

\usepackage[utf8]{inputenc} % input encodings (allow UTF-8 input)
\usepackage[T1]{fontenc} 	% selecting font encodings (use 8-bit T1 fonts)
\usepackage[english]{babel} % multilingual support (English language/hyphenation)

\usepackage[bitstream-charter]{mathdesign}
\usepackage{geometry} 		% flexible and complete interface to document dimensions
\usepackage{mathtools} 		% math package (fixes various deficiencies of amsmath)
\usepackage{float} 			% floating objects such as figures and tables
\usepackage{graphicx} 		% enhanced support for graphics
\usepackage{tabularx} 		% tabulars with adjustable-width columns
\usepackage{booktabs} 		% professional-quality tables
\usepackage{color, xcolor} 	% foreground and background colour management
\definecolor{Rcolor}{HTML}{E99595}
\definecolor{Gcolor}{HTML}{C5E0B4}
\definecolor{Bcolor}{HTML}{9DC3E6}
\definecolor{Ycolor}{HTML}{FFE699}
\definecolor{Vcolor}{HTML}{CFB4E0}
\usepackage{pdfpages} 		% inclusion of external multi-page PDF documents
\usepackage{extarrows} 		% extra arrows beyond those provided in amsmath
\usepackage{multirow} 		% create tabular cells spanning multiple rows
\usepackage{multicol} 		% intermix single and multiple columns
\usepackage{subcaption} 	% support for sub-captions (captions for subfigures)
\usepackage{enumitem} 		% control layout of itemize, enumerate, description
\usepackage{xspace} 		% define commands that appear not to eat spaces
\usepackage{stackrel} 		% enhancement to the \stackrel command
\usepackage{tikz} 			% create PostScript and PDF graphics
\usetikzlibrary{arrows, shapes.geometric, arrows.meta, shapes, decorations.pathreplacing, fit, patterns, patterns.meta, positioning, calc, fadings}
\usepackage{braket} 		% Dirac bra-ket notation
\usepackage{bm} 			% access bold symbols in maths mode
\usepackage{tensor} 		% typeset tensors (tensor-style super- and subscripts)
\usepackage{slashed} 		% slash through characters (Feynman slash notation)
\usepackage{siunitx} 		% SI units package (typesetting values with units)
\usepackage{lastpage} 		% reference last page
\usepackage{cite} 			% improved citation handling
\usepackage[normalem]{ulem} % package for underlining
\usepackage{fontawesome} 	% access to web-related icons
\usepackage{tocloft} 		% control over the typography of the TOC, LOF, and LOT
\usepackage{titlesec} 		% interface to sectioning commands (various title styles)
\usepackage{doi} 			% create correct hyperlinks for DOI numbers
\usepackage{hyperref} 		% hypertext links (handel cross-referencing commands)
\usepackage[most]{tcolorbox} 					% coloured and framed text boxes
\usepackage[nameinlink, capitalize]{cleveref} 	% intelligent cross-referencing
\usepackage[nottoc, notlot, notlof]{tocbibind} 	% add references/index/contents to TOC
\usepackage[ruled, vlined]{algorithm2e} 		% floating algorithm environment
\usepackage{makecell}
\usepackage[makeroom]{cancel}
\usepackage{feynmf}

\makeatletter
\def\BState{\State\hskip-\ALG@thistlm}
\makeatother

\makeatletter
\@ifundefined{pdfoutput}{}{\DeclareGraphicsRule{*}{mps}{*}{}}
\makeatother

\makeatletter
\DeclareRobustCommand*{\bfseries}{%
   \not@math@alphabet\bfseries\mathbf
   \fontseries\bfdefault\selectfont
   \boldmath
}
\makeatother

\fancypagestyle{SPstyle}{
\fancyhf{}
\lhead{\colorbox{scipostblue}{\bf \color{white} ~SciPost Physics Codebases }}
\rhead{{\bf \color{scipostdeepblue} ~Submission }}

\fancyfoot[C]{\textbf{\thepage}}
}

\definecolor{EmeraldGreen}{HTML}{1ea78d}
\definecolor{EnglishRed}{HTML}{b02427}
\hypersetup{
	pdftitle={MadSpace},
	pdfauthor={Heimel et al.},
	colorlinks=true, 			% false: boxed links, true: colored links
	linkcolor={EnglishRed}, 	% color of internal links (sections, pages, etc.)
	citecolor={EmeraldGreen}, 	% color of citation links (links to bibliography)
	urlcolor={EmeraldGreen} 	% color of URL links (external links)
} 
\DeclareSymbolFont{usualmathcal}{OMS}{cmsy}{m}{n}
\DeclareSymbolFontAlphabet{\mathcal}{usualmathcal}

\SetArgSty{textnormal}
\SetKwComment{Comment}{{\small\#}~}{}
\SetCommentSty{mycommfont}

\setitemize{itemsep=0pt, parsep=0pt} 				% adjust itemize environment
\setenumerate{itemsep=0pt, parsep=0pt} 				% adjust enumerate environment
\setitemize{itemsep=2pt,topsep=2pt,parsep=0pt,partopsep=0pt,leftmargin=*}
\setenumerate{itemsep=0pt,topsep=2pt,parsep=0pt,partopsep=0pt,labelindent=3pt,leftmargin=*}
\newlist{todolist}{itemize}{2}
\setlist[todolist]{label=$\square$}
\usepackage{pifont}
\usepackage{amsthm} 		% math package (typesetting theorems using AMS style)
\theoremstyle{definition}

\lstdefinestyle{me7graph}{
    keywordstyle=[0]{\color{purple}},
    keywordstyle=[1]{\color{teal}},
    keywordstyle=[2]{\bfseries},
    basicstyle=\fontfamily{SourceCodePro-TLF}\scriptsize,
}
\lstdefinelanguage{madgraph}{
    keywords=[0]{float, unstack, stable_invariant_nu, sqrt, r_to_x1x2, com_p_in, t_inv_min_max, stable_invariant_nu, two_to_two_particle_scattering_com, mul, two_body_decay, stack, boost_beam, reduce_product, cut_unphysical},
    keywords=[1]{\%0, \%1, \%2, \%3, \%4, \%5, \%6, \%7, \%8, \%9, \%10, \%11, \%12, \%13, \%14, \%15, \%16, \%17, \%18, \%19, \%20, \%21, \%22, \%23, \%24, \%25, \%26, \%27, \%28, \%29, \%30, \%31, \%32, \%33, \%34, \%35, \%36, \%37, \%38, \%39, \%40},
    keywords=[2]{Inputs, Instructions, Outputs},
    alsoletter={\%},
} 
\definecolor{red_cb}{HTML}{e41a1c}
\definecolor{blue_cb}{HTML}{377eb8}
\definecolor{green_cb}{HTML}{4daf4a}
\definecolor{purple_cb}{HTML}{984ea3}
\definecolor{orange_cb}{HTML}{ff7f00}

\definecolor{codebg}{HTML}{F2F4F7}

\newcommand{\ie}{\text{i.e.}\;}

\newcommand{\eqcomma}{\;,} 	% equation comma
\newcommand{\eqperiod}{\;.} 	% equation period

\newcommand{\mwith}{\text{with}}
\newcommand{\mand}{\text{and}}
\newcommand{\mfor}{\text{for}}

\newcommand{\qqquad}{\qquad\quad}

\def\d{\mathrm{d}}

\newcommand\one{\leavevmode\hbox{\small1\normalsize\kern-.33em1}}
\newcommand{\vegas}{\texttt{VEGAS}\xspace}
\newcommand{\madgraph}{\texttt{MadGraph5\_aMC@NLO}\xspace}
\newcommand{\mg}{\texttt{MG5aMC}\xspace}
\newcommand{\mggeneral}{\texttt{MadGraph}\xspace}
\newcommand{\chili}{\texttt{Chili}\xspace}

\newcommand{\alpgen}{\texttt{ALPGEN}\xspace}
\newcommand{\pythia}{\texttt{Pythia}\xspace}
\newcommand{\cudacpp}{\texttt{CUDACPP}\xspace}

\newcommand{\pytorch}{\texttt{PyTorch}\xspace}
\newcommand{\tensorflow}{\texttt{TensorFlow}\xspace}
\newcommand{\numpy}{\texttt{Numpy}\xspace}
\newcommand{\jax}{\texttt{Jax}\xspace}
\newcommand{\sherpa}{\texttt{Sherpa}\xspace}
\newcommand{\herwig}{\texttt{Herwig}\xspace}
\newcommand{\comix}{\texttt{Comix}\xspace}

\newcommand{\pepper}{\texttt{Pepper}\xspace}
\newcommand{\mggpu}{\texttt{MadGraph4GPU}\xspace}
\newcommand{\madspace}{\texttt{MadSpace}\xspace}
\newcommand{\umami}{\texttt{UMAMI}\xspace}
\newcommand{\madnis}{\texttt{MadNIS}\xspace}
\newcommand{\lhapdf}{\texttt{LHAPDF}\xspace}
\newcommand{\python}{\texttt{Python}\xspace}
\newcommand{\rambodiet}{\texttt{RamboOnDiet}\xspace}
\newcommand{\hicom}{\texttt{HICOM}\xspace}
\newcommand{\rambo}{\texttt{Rambo}\xspace}
\newcommand{\fastrambo}{\texttt{FastRambo}\xspace}

\newcommand{\arXiv}[2][]{%
	\ifthenelse{\equal{#1}{}}%
	{\href{http://arxiv.org/abs/#2}{arXiv:#2}}%
	{\href{http://arxiv.org/abs/#2}{arXiv:#2~[#1]}}}

\def\slashchar#1{\setbox0=\hbox{$#1$}           % set a box for #1
   \dimen0=\wd0                                 % and get its size
   \setbox1=\hbox{/} \dimen1=\wd1               % get size of /
   \ifdim\dimen0>\dimen1                        % #1 is bigger
      \rlap{\hbox to \dimen0{\hfil/\hfil}}      % so center / in box
      #1                                        % and print #1
   \else                                        % / is bigger
      \rlap{\hbox to \dimen1{\hfil$#1$\hfil}}   % so center #1
      /                                         % and print /
   \fi}

\newcommand{\tikznode}[2]{%
\ifmmode%
\tikz[remember picture,baseline=(#1.base),inner sep=0pt] \node (#1) {$#2$};%
\else
\tikz[remember picture,baseline=(#1.base),inner sep=0pt] \node (#1) {#2};%
\fi}

\def\mathswitchr#1{\relax\ifmmode{\mathrm{#1}}\else$\mathrm{#1}$\xspace\fi}
\def\mathswitch#1{\relax\ifmmode#1\else$#1$\xspace\fi}

\newcommand{\Pq}{\mathswitch q}
\newcommand{\Pqbar}{\mathswitch{\bar q}}
\newcommand{\PW}{\mathswitchr W}
\newcommand{\PWp}{\mathswitchr {W^+}}

\newcommand{\PZ}{\mathswitchr Z}

\newcommand{\Pg}{\mathswitchr g}
\newcommand{\PH}{\mathswitchr H}

\newcommand{\Pep}{\mathswitchr {e^+}}
\newcommand{\Pem}{\mathswitchr {e^-}}
\newcommand{\Pmup}{\mu^+}
\newcommand{\Pgngm}{\nu_\mu}

\newcommand{\Pu}{\mathswitchr u}

\newcommand{\Pt}{\mathswitchr t}

\newcommand{\Ptbar}{\mathswitchr{\bar t}}
\newcommand{\Pubar}{\mathswitchr{\bar u}}
\newcommand{\Pdbar}{\mathswitchr{\bar d}}

\newcommand{\MZ}{\mathswitch {M_\PZ}}

\graphicspath{{./figs/}}
\begin{document}

\pagestyle{SPstyle}

\begin{center}{\Large \textbf{\color{scipostdeepblue}{MadSpace -- Event Generation for the Era of GPUs and ML}}}
\end{center}

\begin{center}
Theo Heimel\textsuperscript{1},
Olivier Mattelaer\textsuperscript{1},
and Ramon Winterhalder\textsuperscript{2}
\end{center}

\begin{center}
{\bf 1} CP3, Universit\'e catholique de Louvain, Louvain-la-Neuve, Belgium
\\
{\bf 2} TIFLab, Universit\`a degli Studi di Milano \& INFN Sezione di Milano, Italy
\end{center}

\begin{center}
\today
\end{center}

% For convenience during refereeing: line numbers
%\linenumbers

\section*{\color{scipostdeepblue}{Abstract}}
{\bf 
MadSpace is a new modular phase-space and event-generation library written in C++ with native GPU support via CUDA and HIP. It provides a unified compute-graph-based framework for phase-space construction, adaptive and neural importance sampling, and event unweighting. It includes a wide range of mappings, from the standard MadGraph multi-channel phase space to optimized normalizing flows with analytic inverse transformations. All components operate on batches of events and support end-to-end on-device workflows. A high-level Python interface enables seamless integration with machine-learning libraries such as PyTorch.
}

% TODO: include a table of contents (optional)
\vspace{10pt}
\noindent\rule{\textwidth}{1pt}
\tableofcontents
\noindent\rule{\textwidth}{1pt}
\vspace{10pt}

\clearpage
%%%%%%%%%%%%%%%%%%%%%%%%%%%%%%%%%%%%%%%%%%%%%%%%%%%
\section{Introduction}
\label{sec:intro}

One of the cornerstones of precision collider physics is the ability to produce first-principles theoretical predictions that can be directly compared to measured scattering events~\cite{Campbell:2022qmc}. These predictions are generated by event generators such as \pythia~\cite{Bierlich:2022pfr}, \sherpa~\cite{Sherpa:2024mfk}, \herwig~\cite{Bellm:2025pcw}, and \madgraph~\cite{Alwall:2014hca,Frederix:2018nkq} (shortened to \mg hereafter), which build on quantum field theory and are combined in a modular simulation chain. 
However, with the increasing luminosity and data complexity of the upcoming HL-LHC program, there is a growing risk that theoretical simulations will become the bottleneck in the experimental analysis pipeline. Improving simulation speed, scalability, and modularity is therefore crucial to fully exploit the physics potential of future collider runs.

One promising direction to address these challenges is the use of modern machine learning (ML)~\cite{Butter:2022rso,Plehn:2022ftl}, which is becoming a core component of the LHC simulation and analysis pipeline. Neural networks have successfully been applied to speed up expensive scattering amplitude evaluations~\cite{Bishara:2019iwh,Badger:2020uow,Aylett-Bullock:2021hmo,Maitre:2021uaa,Danziger:2021eeg,Winterhalder:2021ngy,Badger:2022hwf,Maitre:2023dqz,Spinner:2024hjm,Brehmer:2024yqw,Breso-Pla:2024pda,Bahl:2024gyt,Janssen:2023ahv,Herrmann:2025nnz,Bahl:2025xvx,Villadamigo:2025our,Beccatini:2025tpk,Bahl:2026qaf,Bahl:2026jvt}, and improve phase-space sampling~\cite{Bendavid:2017zhk,Klimek:2018mza,Chen:2020nfb,Gao:2020vdv,Bothmann:2020ywa,Gao:2020zvv,Heimel:2022wyj,Heimel:2023ngj,Heimel:2024wph,Janssen:2025zke,Bothmann:2025lwg}. 
ML-driven approaches have also demonstrated substantial gains in other stages of the simulation and analysis chain -- including parton showers, hadronisation, detector simulation, reconstruction, and analysis -- for which we refer to the HEP-ML Living Review~\cite{Feickert:2021ajf,hepmllivingreview}.

Despite these advances, many computational bottlenecks persist. Phase-space integration and event-generation efficiency remain central challenges in modern HEP simulation pipelines~\cite{Brass:2018xbv,Lepage:2020tgj,Heraiz:2026bok,Isaacson:2023iui,Bothmann:2023siu} and are active fields of research within \mg \cite{Maltoni:2002qb,Mattelaer:2021xdr,Frederix:2026ejl,Hagiwara:2026lul}.
These challenges are exacerbated by the performance limitations of traditional CPU-based infrastructures and by the architectural rigidity of legacy simulation code.
As a result, hardware acceleration and massively parallel architectures have emerged as a complementary and increasingly important direction.
Early pioneering work has demonstrated the feasibility of porting helicity amplitude calculations to GPUs~\cite{Hagiwara:2010oca,Hagiwara:2010ujr,Giele:2010ks,Hagiwara:2013oka}.
More recently, the \pepper framework~\cite{Bothmann:2023gew} provides a first end-to-end GPU-based event generator, including fast amplitude evaluations~\cite{Bothmann:2021nch}, and integration routines based on \chili~\cite{Bothmann:2023siu}. Related efforts have explored \tensorflow-based GPU acceleration~\cite{Carrazza:2020rdn,Carrazza:2020qwu,Carrazza:2021gpx}.
Similarly, developments within the \mggpu effort have demonstrated significant speed\-ups for matrix-element calculations on GPUs~\cite{Hageboeck:2023blb,Wettersten:2025hrb,Valassi:2025xfn,Hagebock:2025jyk} and event reweighting~\cite{Roiser:2025bsk}, but have so far solely relied on CPU-based sampling and integration methods. A fully featured and modular GPU-based phase-space generator --- supporting standard techniques such as multi-channel integration and \rambo-style mappings~\cite{Rambo, vanHameren:2003pr,RamboDiet} --- is missing in the \mggeneral ecosystem. 

In this paper, we present \madspace, a new modular and hardware-optimized phase-space and event-generation library written in \texttt{C++} including CUDA and HIP kernels, designed from the ground up to run efficiently on both CPUs and GPUs.
It provides high-performance implementations of common Feynman-diagram-based mappings and new \rambo-like schemes, including a GPU-tuned \fastrambo variant. Adaptive sampling routines, such as \vegas~\cite{Lepage:1977sw,Lepage:1980dq,Lepage:2020tgj} and normalizing flows, are supported natively on GPUs, alongside a lightweight parton distribution interface. Combined with support for inverse mappings and a streamlined \python API, \madspace offers the essential missing link towards a fully GPU-enhanced simulation pipeline. Its modular design enables easy interfacing with machine learning libraries like the \madnis framework~\cite{Heimel:2022wyj,Heimel:2023ngj}, and opens the door for efficient inference, surrogate modeling, and differentiable programming~\cite{Heinrich:2022xfa,Nachman:2022jbj,MODE:2022znx,Kagan:2023gxz,Aehle:2024ezu,Heimel:2024wph}.

The upcoming major release of MadGraph7, which is already publicly available as an alpha release~\cite{MadGraph7}, incorporates \madspace~\cite{MadSpace} as a core component.
Its initial deployment targets LO computations, with NLO functionality scheduled for later releases.

%\clearpage
%%%%%%%%%%%%%%%%%%%%%%%%%%%%%%%%%%%%%%%%%%%%%%%%%%%
\section{Phase-space mappings}
\label{sec:ps_mappings}

At leading order (LO), the evaluation of cross sections and the generation of unweighted events reduce to the numerical computation of integrals of the form
\begin{equation}
    I = \int_{\Phi} \d x \, f(x)\eqcomma
    \label{eq:ps_integral}
\end{equation}
where $x$ denotes a $d$-dimensional phase-space point and $f(x)$ is the fully differential cross section, given by the squared matrix element times flux factors, parton distribution functions (PDFs), and selection cuts. 
In general, an analytical evaluation of this integral is not feasible and one has to use Monte Carlo (MC) techniques. 
In the MC approach, we typically start by introducing an invertible mapping 
\begin{equation}
r \in U = [0,1]^d \quad 
\xleftrightarrow[\quad \leftarrow G^{-1}(x)\quad]{G(r)\rightarrow} 
\quad x\in\Phi\subseteq\mathbb{R}^d\eqcomma
\label{eq:ps_mapping}
\end{equation}
from a unit-hypercube onto the physical phase space.
This induces a normalized sampling density given by the Jacobian determinant
\begin{equation}
    g(x) = \left|\frac{\partial G^{-1}(x)}{\partial x} \right| \qquad \mwith
    \qquad \int_{\Phi} \d x \, g(x) = 1 \eqcomma
    \label{eq:ps_mapping_basis}
\end{equation}
such that the integral can be rewritten as
\begin{equation}
    I
    = \int_{\Phi} \d x \, g(x)\,\frac{f(x)}{g(x)}
    = \int_{U} \d r \, \left.\frac{f(x)}{g(x)}\right|_{x=G(r)}\eqperiod
    \label{eq:ps_importance_sampling}
\end{equation}
While the integral is unchanged under this reparametrization, the variance of the new integrand is given by
\begin{equation}
\begin{split}
  \text{Var}_g\left[\frac{f}{g}\right]
  = \int_\Phi \d x\, g(x) \left(\frac{f(x)}{g(x)}-I\right)^2\eqcomma
\end{split}
\label{eq:def_sigma}
\end{equation}
which is minimized for the ideal choice $g(x)=f(x)/I$.
Efficient integration and event generation therefore require a sampling density $g(x)$ that approximates the target distribution $f(x)$ as closely as possible over the entire phase space.

%%%%%%%%%%%%%%%%%%%%%%%%%%%%%%%%%%%%%%%%%%%%%%%%%%%
\subsection{Multi-channel integration}
\label{sec:ps_multichannel}

In realistic collider applications, the integrand is a combination of very different structures, including narrow resonances, soft and collinear enhancements, and phase-space cuts. 
No single global mapping $G(x)$ can efficiently resolve all of these features simultaneously over the full phase space. 
This motivates a decomposition of the sampling density into several complementary channels, each tailored to a subset of the dominant structures. 
Two conceptually different realizations of this idea coexist in the literature and in practical codes, and the distinction is essential for understanding the multi-channel strategy employed in \madspace.

%%===================================
\subsubsection{Global vs.\ local multi-channeling}

In the standard multi-channel approach~\cite{Kleiss:1994qy, Weinzierl:2000wd}, we start by introducing several channel mappings $G_i:U_i=[0,1]^d \to \Phi$ denoted as $r \to x = G_i(r)$, to obtain individual densities
\begin{equation}
   g_i(x)=\left\vert\frac{\partial G^{-1}_i(x)}{\partial x}\right\vert
   \qquad \mwith \qquad \int_\Phi \d x\,g_i(x)=1
   \qquad \mfor \qquad i=1,\dots,M\eqcomma
   \label{eq:channel_densities}
\end{equation}
where $M$ is the total number of channels. 
We can then combine the individual channel densities into a total normalized density
\begin{equation}
    g(x) = \sum_{i=1}^{M} \beta_i \, g_i(x) \qquad \mwith \qquad
    \sum_{i=1}^{M} \beta_i = 1\eqcomma
    \quad \mand \quad
    \beta_i \ge 0 \eqcomma
\end{equation}
with \emph{global}, phase-space independent weights $\beta_i$. 
This allows us to reparametrize Eq.\eqref{eq:ps_integral} into
\begin{equation}
    I=\sum_{i=1}^{M} \beta_i \int_{\Phi}\d x\,g_i(x)\,\frac{f(x)}{g(x)}
    =\sum_{i=1}^{M} \beta_i \int_{U_i}\d r\,\left.\frac{f(x)}{g(x)}\right\vert_{x=G_i(r)}\eqperiod
    \label{eq:multi-channel-standard}
\end{equation}
While each channel generates its events from independent mappings $G_i$, they all evaluate the same integrand or weight $w(x)=f(x)/g(x)$. 
The mixture weights $\beta_i$ are optimized to minimize the total variance~\cite{Kleiss:1994qy, Weinzierl:2000wd}.
This is the strategy traditionally employed in adaptive multi-channel integrators and is also the conceptual basis within \sherpa.

In contrast, \mg follows a different strategy in its single-diagram enhanced (SDE) integration method~\cite{Maltoni:2002qb, Mattelaer:2021xdr}. Here the phase-space decomposition is written as
\begin{equation}
    I = \int_\Phi \d x \, f(x)
    = \sum_{i=1}^{M} \int_\Phi \d x \, \alpha_i(x)\, f(x)
    \qquad \mwith \qquad \sum_{i=1}^{M} \alpha_i(x)=1 \eqcomma
\end{equation}
with \emph{local}, phase-space dependent channel weights $\alpha_i(x)$.
In the most common realization, these weights are constructed from diagram-level information, either from single or sub-am\-pli\-tudes $|{\cal M}_i(x)|^2$~\cite{Maltoni:2002qb}, referred to as SDE strategy~1, or from products of propagator denominators~\cite{Mattelaer:2021xdr}, referred to as SDE strategy~2.
Inserting this into Eq.\eqref{eq:ps_integral}, we can decompose and parameterize the phase-space integral as
\begin{equation}
  I
  = \sum_{i=1}^{M}\int_\Phi\d  x\, \alpha_i(x)\,f(x)
  = \sum_{i=1}^{M}\int_{U_i}\d  r\, \alpha_i(x)\left.\frac{f(x)}{g_i(x)}\right\vert_{x=G_i(r)}\eqperiod
  \label{eq:multi-channel-mg}
\end{equation}
While both formulations are mathematically equivalent and coincide for
\begin{equation}
  \alpha_i(x)=\beta_i\,\frac{g_i(x)}{g(x)}\eqcomma
  \label{eq:multi-channel-unified}
\end{equation}
they lead to rather different algorithmic structures.
To make this explicit, we consider the Monte Carlo estimate of the integral.
In both formulations, the integral of each channel is estimated from $N_i$ uniformly distributed random points $r_{i}\in U_i$, which are mapped onto the phase space with the fixed mapping of that channel,
\begin{equation}
  I \approx \sum_{i=1}^{M} \frac{1}{N_i} \sum_{k=1}^{N_i} w_i(x_{ik})
  \qquad \mwith \qquad
  x_{ik} = G_i(r_{ik})\eqperiod
  \label{eq:multi-channel-mc}
\end{equation}
Each phase-space point is therefore generated by, and attributed to, one specific channel.
The number of samples per channel $N_i$ is in principle arbitrary, but is typically chosen to minimize overall variance in a pure integrator, or to yield the required number of unweighted events proportional to each channel's contribution to the total cross-section in an event generator.
According to Eqs.\eqref{eq:multi-channel-standard} and~\eqref{eq:multi-channel-mg}, the two formulations differ only in the event weights
\begin{equation}
  w_i^\text{std}(x) = \beta_i\,\frac{f(x)}{g(x)}
  \qquad \mand \qquad
  w_i^\text{SDE}(x) = \alpha_i(x)\,\frac{f(x)}{g_i(x)}\eqperiod
  \label{eq:multi-channel-weights}
\end{equation}
In the standard approach with overlapping mappings, the event weight depends on the full mixture $g(x)$, such that every event requires all densities $g_j(x)$ and therefore the inverse direction of every mapping.
Conversely, if the mappings $G_i$ cover disjoint regions of phase space, only a single density is non-zero at each point, and the standard approach also requires only the density of the generating mapping.
The usual choice $N_i=\beta_i N$ corresponds to drawing the channel index with probability $\beta_i$ and reduces Eq.~\eqref{eq:multi-channel-mc} to an average of $f(x)/g(x)$ over points sampled from $g(x)$.
In the SDE approach, $\alpha_i(x)$ is constructed from diagram-level quantities that depend on the momenta alone, such that only the density of the generating mapping is required.
It follows from the forward map using the inverse function rule
\begin{equation}
    g_i(G_i(r))=\left\vert\frac{\partial G_i(r)}{\partial r}\right\vert^{-1}\eqperiod
\end{equation}
The relevant distinction is therefore not whether the channel weights are local or global, but whether the event weight refers to the densities of the other mappings.
In \mg, each channel has its own \vegas grid, and the number of points $N_i$ is adapted to the contribution of the channel to the total cross section.
In \madspace, we implement all phase-space transformations as fully invertible maps and thus retain the flexibility to realize both multi-channel strategies within a common framework.

The construction of the channel mappings $G_i$ is guided by physics insight into the analytic structure of scattering amplitudes, such as resonant propagators, soft and collinear limits, and small momentum-transfer regions. These physics-motivated parametrizations provide a robust starting point, which can subsequently be refined by adaptive techniques such as \vegas~\cite{Lepage:1977sw,Lepage:1980dq,Lepage:2020tgj} or by learned importance sampling as in \madnis~\cite{Heimel:2022wyj,Heimel:2023ngj,Heimel:2024wph}.

In the following, we describe the set of analytic phase-space mappings implemented in \madspace that are most relevant for the present study.
They are primarily based on a recursive decomposition into decay and scattering blocks inspired by Feynman-diagram topologies, and we include in addition a fast variant of \rambo-like algorithms~\cite{Rambo,vanHameren:2003pr, RamboDiet}. 
An alternative parametrization of the final-state kinematics in terms of transverse momenta, rapidities, and azimuthal angles, following the \chili~\cite{Bothmann:2023siu} approach is also available in \madspace.

%%%%%%%%%%%%%%%%%%%%%%%%%%%%%%%%%%%%%%%%%%%%%%%%%%%
\subsection{Recursive phase-space decomposition}
\label{sec:fd_mappings}

The channel mappings employed in \madspace are based on the recursive decomposition of the tree-level $n$-body phase space, organized as a sequence of elementary scattering and decay building blocks. By aligning the integration variables with the physical quantities that describe the dominant structure of the matrix element, we can construct mappings that locally flatten the dominant variations of the integrand.

For a given $2\to n$ process and a chosen diagram topology, we introduce a set of time-like invariants $s_i$ associated with internal propagators, a set of space-like momentum transfers $t_i$ associated with $t$-channel exchanges, and a sequence of two- and three-particle phase-space building blocks.
The resulting $n$-body phase space can be written in the generic form
\begin{equation}
\begin{split}
\int \d \Phi_n(x)
&=
\left[ \prod_{i=1}^{n-2-\gamma} \int \d s_i \right]\times
\left[ \prod_{j=1}^{\kappa-\beta} \int \d \Phi^{(\phi,t)}_{2,j}(x) \right]\times
\left[ \prod_{k=1}^{\beta} \int \d \Phi^{(\tilde s,t)}_{2,k}(x) \right] \\
&\phantom{=}\times
\left[ \prod_{l=1}^{\gamma} \int \d \Phi_{3,l}(x) \right]\times
\left[ \prod_{m=1}^{n-\kappa-1-2\gamma} \int \d \Phi^{(\phi,\theta)}_{2,m}(x) \right]
\eqperiod
\end{split}
\label{eq:ps_decomposition}
\end{equation}
Here $\d\Phi_2$ denotes a two-particle phase space and $\d\Phi_3$ a genuine three-particle phase space. Superscripts indicate the variables used to parametrize the corresponding phase-space element, rather than a specific subprocess interpretation. Different superscripts on $\d\Phi_2$ therefore correspond to equivalent parametrizations of the same two-particle phase space.
Throughout this section, we omit the conventional overall factors of $(2\pi)^{4-3n}$ in the definition of $\d\Phi_n$ for readability. These factors are, of course, included in the actual implementation to obtain correctly normalized cross sections.

The integer $\kappa$ denotes the number of space-like momentum transfers in the chosen topology, $\beta$ the number of two-particle phase-space blocks parametrized in terms of a pair of invariants $(\tilde s,t)$, and $\gamma$ the number of three-particle decay blocks. 
The remaining two-particle phase-space factors correspond to alternative parametrizations that are naturally associated with scattering- or decay-like kinematic configurations. For $\kappa \ge 2$, not all time-like invariants $s_i$ are associated with physical propagators.

Counting the independent integration variables of each building block, we can organize the $(3n-4)$-dimensional phase space as:
\begin{itemize}
    \item $d_{\text{inv}} = n-2-\gamma$ degrees of freedom from time-like invariants $s_i$;
    \item $d_{\text{2p}} = 2\kappa$ degrees of freedom from two-particle phase-space blocks that require incoming reference momenta, \ie parametrizations naturally associated with scattering kinematics;
    \item $d_{\text{decay}} = 2(n-\kappa-1)+\gamma$ degrees of freedom from two- and three-particle decay blocks.
\end{itemize}
Together with the usual two PDF convolution variables, this yields a total of $3n-2$ integration dimensions. For each of these elementary phase-space factors one can define an explicit and invertible mapping between the physical variables and the unit hypercube, as is commonly done in multi-purpose event generators~\cite{Sjostrand:2014zea, Alwall:2014hca, Sherpa:2019gpd}.

As the construction of a full $2\to n$ phase-space mapping proceeds recursively, we organize the presentation as follows:
We first introduce a set of generic and recurring analytic phase-space mappings, which define invertible transformations between the unit hypercube and physical phase space and are reused across multiple building blocks to locally flatten dominant variations of the integrand. We then describe how these mappings are employed in the PDF convolution and in
the parametrizations of the two- and three-particle phase-space elements, corresponding to angular, $t$-channel, and invariant representations of the two-particle phase space, as well as the genuinely three-particle phase space.

%%===================================
\subsubsection{Analytic phase-space mappings}

Analytic phase-space mappings define invertible transformations from uniformly distributed random numbers $r\in[0,1]$ to physical phase-space
variables $x\in\Phi$, together with the associated Jacobian densities. They are used to reflect dominant structures of the integrand, such as resonances, thresholds, soft or collinear enhancements, or phase-space
boundaries, and are reused across multiple phase-space building blocks to locally flatten the dominant variations of the integrand. For resonant structures of the generic form
\begin{equation}
    |\mathcal{M}|^2 \;\propto\; \frac{1}{(x-m^2)^2+m^2\Gamma^2}\eqcomma
\end{equation}
we use the  mapping $x = G(r)$ such that
\begin{equation}
    \int_{x_{\min}}^{x_{\max}} \d x
    = \int_0^1 \frac{\d r}{g(x(r),x_{\min},x_{\max})}
    \qquad \mwith \qquad
    g(x)=\left|\frac{\partial G^{-1}(x)}{\partial x}\right|\eqperiod
\end{equation}
For a finite-width resonance we employ a Breit--Wigner mapping,
\begin{equation}
\begin{split}
    G_{\text{BW}}(r,m^2,x_{\min},x_{\max})
    &= m\Gamma \tan\!\left[u_1+(u_2-u_1)r\right] + m^2 \eqcomma \\
    g_{\text{BW}}(x,m^2,x_{\min},x_{\max})
    &= \frac{m\Gamma}
       {(u_2-u_1)\left[(x-m^2)^2+m^2\Gamma^2\right]}\eqcomma
\end{split}
\label{eq:prop_massive}
\end{equation}
with
\begin{equation}
u_{1/2} = \arctan\left(\frac{x_\mathrm{min/max}-m^2}{m\Gamma}\right)\eqperiod
\end{equation}
For non-resonant or effectively stable structures ($\Gamma=0$) we use a power-law mapping
\begin{equation}
\begin{split}
    G_{\nu}(r, m^2, x_{\min}, x_{\max})
    &= \left[
        r (x_{\max}-m^2)^{1-\nu}
        + (1-r)(x_{\min}-m^2)^{1-\nu}
       \right]^{\frac{1}{1-\nu}} + m^2 \eqcomma \\
    g_{\nu}(x, m^2,x_{\min},x_{\max})
    &= \frac{1-\nu}
       {\left[(x_{\max}-m^2)^{1-\nu}-(x_{\min}-m^2)^{1-\nu}\right](x-m^2)^{\nu}}\eqcomma
\end{split}
\label{eq:prop_stable_nu}
\end{equation}
valid only for $\nu\neq1$. The special case $\nu\to1$ is obtained as a smooth limit and yields a logarithmic mapping
\begin{equation}
\begin{split}
    G_{\nu=1}(r,m^2,x_{\min},x_{\max})
    &= \exp\!\left[
        r\,\log(x_{\max}-m^2)
        + (1-r)\log(x_{\min}-m^2)
       \right] + m^2 \eqcomma \\
    g_{\nu=1}(x,m^2,x_{\min},x_{\max})
    &= \frac{1}{(x-m^2)\,
       \left[\log(x_{\max}-m^2)-\log(x_{\min}-m^2)\right]}\eqperiod
\end{split}
\label{eq:prop_stable_nu1}
\end{equation}
The mappings in Eq.\eqref{eq:prop_stable_nu} and its logarithmic limit Eq.\eqref{eq:prop_stable_nu1} are strictly well defined only for $x_{\min}-m^2>0$, \ie when the lower integration boundary lies above the pole
position. For massless structures, or more generally when $m^2=0$ and $x_{\min}=0$, this condition is not satisfied. We therefore introduce a small auxiliary negative mass parameter $m^2=-a$ with $0<a\ll1$ in the mapping only.
This stabilizes the logarithmic and power-law transformations near the phase-space boundary, while leaving the physical matrix element and the
integrand $f(x)$ unchanged. In practice, the exponent $\nu$ can be tuned to optimize variance reduction and unweighting efficiency. The naive expectation $\nu=2$ is not necessarily optimal; in our implementation we use $\nu=0.8$ as a default choice.

For phase-space variables that are not associated with any pronounced structure in the integrand, it is often sufficient to sample the variable uniformly within
its allowed interval. This corresponds to the choice $\nu=0$, for which the mapping reduces to
\begin{equation}
\begin{split}
    G_{\text{flat}}(r,x_{\min},x_{\max})
    &= x_{\min} + (x_{\max}-x_{\min})\,r \eqcomma \\
    g_{\text{flat}}(x,x_{\min},x_{\max})
    &= \frac{1}{x_{\max}-x_{\min}}\eqperiod
\end{split}
\label{eq:prop_flat}
\end{equation}
These analytic mappings provide a robust baseline and are complemented by numerical and adaptive techniques, such as \vegas-style grid adaptation and neural
importance sampling with \madnis, which further refine the sampling density and improve integrand flattening in high-dimensional phase spaces.

%%===================================
\subsubsection{PDF convolutions}

The initial-state kinematics are described by the parton momentum fractions $x_1$ and $x_2$, which enter the phase-space measure through the convolution with PDF and determine the squared partonic center-of-mass energy
$\hat{s}=x_1 x_2 s$. Using $\hat{s}$ as integration variable, we can write the PDF convolution as
\begin{equation}
\int_{0}^{1} \d x_1 \d x_2\,
\Theta\!\left(\hat{s}-\hat{s}_\text{min}\right)
= \int_{\hat{s}_\text{min}}^{s} \d \hat{s}
  \int_{\hat{s}/s}^{1} \frac{\d x_1}{x_1s}\eqcomma
\label{eq:pdf_conv}
\end{equation}
and where $\hat{s}_\text{min}$ is fixed by final-state masses and analysis cuts. The second integration variable can be chosen as $x_1$, with $x_2=\hat{s}/(x_1 s)$. After fixing the partonic invariant $\hat{s}$, we parametrize the momentum fractions as
\begin{equation}
x_1 = \tau^r
\qquad \mand \qquad
x_2 = \tau^{1-r}
\qquad \mwith \qquad \tau=\frac{\hat{s}}{s}\eqperiod
\label{eq:pdf_sampling}
\end{equation}
This construction corresponds to logarithmic sampling of $x_1$ at fixed $\hat{s}$ yielding
\begin{equation}
\d x_1 = x_1\,\log\tau \;\d r \eqcomma
\end{equation}
so that the $1/x_1$ dependence of the convolution measure in Eq.\eqref{eq:pdf_conv} is canceled by the sampling density. 
The invariant $\hat{s}$ itself is sampled using an analytic mapping introduced above. If $\hat{s}$ coincides with the first time-like invariant of the chosen channel and is associated with an $s$-channel resonant propagator -- excluding topologies with additional $t$-channel propagators -- we employ a Breit--Wigner mapping $\hat{s}=G_{\text{BW}}(r)$ to directly resolve the resonance peak. Otherwise, we use a power-law mapping $\hat{s}=G_{\nu}(r)$; choosing $\nu\simeq 1$ absorbs the $1/\hat{s}$ behavior of the flux factor appearing in the partonic cross section. 
We note that the PDF convolution and $\hat{s}$-based parametrization described here are specific to hadron-initiated processes.
For lepton colliders with ISR or effective lepton structure functions, the natural integration variables are lepton energy fractions rather than $\hat{s}$, which will be treated in a future implementation.

%%===================================
\subsubsection{Two-particle phase space in \texorpdfstring{$(\phi,\theta)$}{(phi,theta)} variables}
\label{sec:one_to_two}

%------------------------------------------
\begin{figure}[b!]
    \centering
    \includegraphics[width=0.30\textwidth]{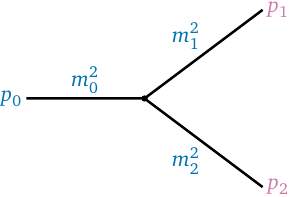}
    \hspace{3mm}
    \includegraphics[width=0.30\textwidth]{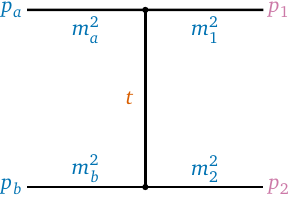}
    \hspace{3mm}
    \includegraphics[width=0.30\textwidth]{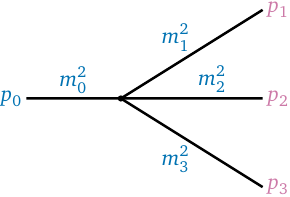}
    \caption{From left to right: The kinematics for a basic $1\to2$, $2\to2$, and $1\to3$ building block. The input variables (blue), sampled invariants (orange), and constructed momenta (purple) are color highlighted. }
    \label{fig:diagrams_simple_blocks}
\end{figure}
%------------------------------------------

We first consider an elementary two-particle phase-space element, as shown in Fig.~\ref{fig:diagrams_simple_blocks}, corresponding to a parent momentum $p_0$ decaying into two final-state momenta $p_1$ and $p_2$, with $p_0 = p_1 + p_2$.
The invariant mass $m_0^2 = p_0^2$ is fixed by the upstream phase-space construction, and the daughter masses $m_{1,2}$ are given. In the rest frame of $p_0$, the energies and magnitudes of the three-momenta are fixed by two-body kinematics
\begin{equation}
E_{1/2} = \frac{m_0^2 \pm (m_1^2 - m_2^2)}{2m_0} \qquad \mand \qquad
|\vec p_{1/2}|=\frac{\sqrt{\lambda(m_0^2,m_1^2,m_2^2)}}{2m_0}\eqcomma
\end{equation}
where $\lambda(x,y,z)=(x-y-z)^2-4yz$ denotes the Källén function. 
The remaining degrees of freedom are purely angular.
We therefore choose the polar and azimuthal angles $(\theta,\phi)$ of $\vec p_1$ in the
rest frame of $p_0$ as integration variables and sample them uniformly
\begin{equation}
\phi = 2\pi r_\phi \qquad \mand \qquad \cos\theta = 2 r_\theta - 1 \eqperiod
\end{equation}
The momenta $p_{1,2}$ are then constructed from these angles and boosted back to the frame defined by $p_0$.
The corresponding phase-space measure can be written directly in terms of the physical variables as
\begin{equation}
\begin{split}
\int \d \Phi^{(\phi,\theta)}_2(x)
&=
\frac{\sqrt{\lambda(m_0^2,m_1^2,m_2^2)}}{8m_0^2}
\int_{0}^{2\pi} \d\phi
\int_{-1}^{1} \d\!\cos\theta\eqperiod
\end{split}
\label{eq:ps2_phi_theta}
\end{equation}

%%===================================
\subsubsection{Two-particle phase space in \texorpdfstring{$(\phi,t)$}{(phi,t)} variables}
\label{sec:two_to_two}

We now reparametrize the two-particle phase space introduced above in terms of the Mandelstam invariant $t$ instead of the polar angle $\theta$. This choice is particularly convenient for kinematic configurations that permit an interpretation in terms of a $2\to2$ scattering subprocess, where the dominant singular behaviour of the matrix element is typically associated with a $t$-channel propagator.

We therefore consider a generic two-particle final state with
$p_a+p_b=p_1+p_2$ and total momentum $p_0=p_a+p_b$.
The scattering center-of-mass energy $m_0^2=p_0^2$ and the external masses $m_{i}$ are fixed or provided by other components of the phase-space construction.
The phase space is parametrized in terms of the azimuthal angle $\phi$ and the Mandelstam variable $t=(p_a-p_1)^2<0$.
Introducing the shorthands
\begin{equation}
\Delta_{ij} = m_i^2 - m_j^2 \qqquad \mand \qqquad
\Lambda_{ij} = \lambda(m_0^2,m_i^2,m_j^2)\eqcomma
\end{equation}
the invariant $t$ depends linearly on $\cos\theta$ as
\begin{equation}
t = m_1^2+m_a^2
 -\frac{(m_0^2+\Delta_{12})\,(m_0^2+\Delta_{ab})
        -\sqrt{\Lambda_{12}\Lambda_{ab}}\cos\theta}
       {2m_0^2}\eqperiod
\label{eq:t_invariant_theta}
\end{equation}
The integration limits $t_{\min}$ and $t_{\max}$ follow from $-1\le\cos\theta\le1$. In practice, it is convenient to work with the positive quantity $|t|=-t$, whose allowed range is given by $|t|\in[-t_{\max},-t_{\min}]$. The phase space is therefore parametrized by the variables $(|t|,\phi)$ and we sample the variables according to
\begin{equation}
    \phi = 2\pi r_\phi \qquad \mand \qquad
    |t| = G_{\nu}(r_t, 0, -t_{\max}, -t_{\min}) \eqcomma
    \label{eq:2_to_2_mapping}
\end{equation}
where $G_{\nu}$ is the generic power-law invariant
mapping defined in Eq.\eqref{eq:prop_stable_nu}. 
The momenta $p_{1,2}$ are constructed from the angles
$(\theta=\theta(t),\phi)$ in the scattering center-of-mass frame and boosted back to the frame defined by $p_0$.
The corresponding phase-space measure can be written directly in terms of the physical variables as
\begin{equation}
\begin{split}
\int \d \Phi^{(\phi,t)}_2(x)
&=
\frac{1}{4\sqrt{\lambda(m_0^2,m_a^2,m_b^2)}}
\int_{0}^{2\pi} \d\phi
\int_{-t_{\max}}^{-t_{\min}}
\d |t|
\eqperiod
\end{split}
\label{eq:ps2_phi_t}
\end{equation}
%

%%===================================
\subsubsection{Two-particle phase space in \texorpdfstring{$(\tilde s,t)$}{(s,t)} variables}
\label{sec:two_to_three}

In addition to the angular and $(\phi,t)$ parametrizations discussed above, the two-particle phase space further admits a double-invariant representation in which both angular variables are replaced by Lorentz invariants. Starting from the $(\phi,t)$ parametrization, this corresponds to replacing
the azimuthal angle $\phi$ by an additional time-like invariant $\tilde s$. Such a parametrization was first introduced in the work of Ref.~\cite{Byckling:1969luw}, and studied in detail in Ref.~\cite{Frederix:2024uvy}. We denote the resulting double-invariant two-particle phase-space element by $\d\Phi^{(\tilde s,t)}_2$.

Unlike the replacement of the polar angle by the momentum transfer $t$, the substitution of the azimuthal angle by a time-like invariant does not, by itself, provide a complete parametrization of the outgoing momenta. While fixing the pair of invariants $(s,t)$ uniquely determines the kinematics at the level of Lorentz invariants, the explicit construction of the final-state momenta requires additional directional information. In practice, this information is supplied by a recoil momentum, which serves to define the scattering plane and hence the azimuthal orientation of the system.
As a consequence, the $\d\Phi^{(\tilde s,t)}_2$ building block naturally appears as part of a larger recursive phase-space construction, in which one of the participating momenta represents a composite subsystem that is resolved only at a later stage.

\begin{figure}[t!]
    \centering
    \begin{minipage}[c]{0.45\textwidth}
        \centering
        \includegraphics[height=5.5cm]{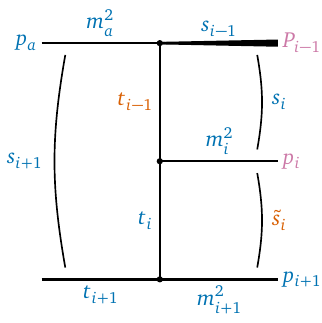}
    \end{minipage}
    \hspace{1mm}
    \begin{minipage}[c]{0.45\textwidth}
        \centering
        \includegraphics[height=4.5cm]{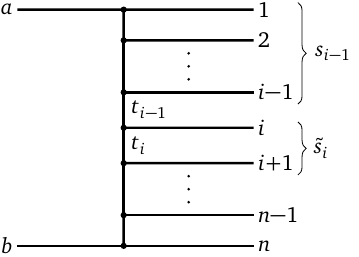}
    \end{minipage}
    \caption{Left: Basic building block with color highlighted input variables (blue), sampled invariants (orange) and constructed momenta (purple). Right: General $2\to n$ process with the general definitions of the variables $\hat{s}_{i}$, $t_i$, and $s_{i}$.}
    \label{fig:diagram_channel_23}
\end{figure}

This situation is illustrated in the left panel of Fig.~\ref{fig:diagram_channel_23}. At this stage of the recursive construction, the system consists of an on-shell momentum $p_{i+1}$ with mass $m_{i+1}$ and a composite cluster with total momentum $P_i = p_1 + \dots + p_i$ and invariant mass $s_i = P_i^2$.
The $\d\Phi^{(\tilde s,t)}_2$ building block then resolves this cluster by peeling off a single on-shell momentum $p_i$ of mass $m_i$, leaving a reduced downstream cluster $P_{i-1} = p_1 + \dots + p_{i-1}$ with invariant mass $s_{i-1}=P_{i-1}^2$. Note that both $s_i$ and $s_{i-1}$ have been generated previously by the time--like-invariants building block in Eq.\eqref{eq:ps_decomposition}.
The momenta $p_i$ and $P_{i-1}$ are constructed using the invariants $(\tilde s_{i}, t_{i-1})$ together with the recoil provided by $p_{i+1}$. This procedure can be repeated recursively until the remaining cluster has been fully resolved. The right panel of Fig.~\ref{fig:diagram_channel_23} shows how such building blocks are embedded into a general $2\to n$ phase-space topology.

%------------------------------------------------------
%------------------------------------------------------

Mathematically, the $\d\Phi^{(\tilde s,t)}_2$ block trades the angular parametrization of a two-particle decay for a double-invariant representation, in which the polar and azimuthal angles are replaced by a space-like momentum transfer and an additional time-like invariant. This leads to
\begin{equation}
\begin{split}
\int \d \Phi_2^{(\tilde s,t)}(x)
&=\int_{\tilde{s}^\text{min}_{i}}^{\tilde{s}^\text{max}_{i}}\d \tilde{s}_{i}
  \int_{-t_{i-1}^{\max}}^{-t_{i-1}^{\min}}\d |t_{i-1}|\,
  \frac{1}{8\sqrt{-\Delta_4(p_a,q_{i-1},q_i,q_{i+1})}}\eqcomma
\label{eq:byckling}
\end{split}
\end{equation} 
where we have introduced the shorthand notations
\begin{equation}
q_{i}=p_a-P_{i} \qqquad q^2_i=t_i \qqquad
P_{i}=p_1+\cdots+p_i \qqquad P^2_i=s_i \eqperiod
\end{equation}
The Gram determinant appearing in Eq.\eqref{eq:byckling} is defined as
\begin{equation}
\Delta_4=\frac{1}{16}
\begin{vmatrix}
2m_a^2 & m_a^2+t_{i-1}-s_{i-1} & m_a^2+t_i-s_i & m_a^2+t_{i+1}-s_{i+1} \\[0.1cm]
& 2t_{i-1} & t_{i-1}+t_i-m_i^2 & t_{i-1}+t_{i+1}-\tilde{s}_i \\[0.1cm]
&      & 2t_i          & t_i+t_{i+1}-m_{i+1}^2 \\[0.1cm]
&      &               & 2t_{i+1} 
\end{vmatrix}\eqperiod
\end{equation}
The polar angle $\cos\theta_{i-1}$ of the momentum $P_{i-1}$ in the $P_i$ rest frame is linearly connected to the space-like invariant $t_{i-1}$ as 
\begin{equation}
t_{i-1}=s_{i-1}+m_a^2-\frac{(s_i+s_{i-1}-m_i^2)(s_i+m_a^2-t_i)-\sqrt{\lambda(s_i,s_{i-1},m_i^2)\lambda(s_i,t_i,m_a^2)}\cos\theta_{i-1}}{2s_i}\eqperiod
\label{eq:t_invariant_theta_23}
\end{equation}
Moreover, the azimuthal angle $\phi_{i-1}$ is related to the time-like invariant $\tilde{s}_i$ by
\begin{equation}
\tilde{s}_i=s_{i+1}+s_{i-1}+\frac{8V+8\cos\phi_{i-1}\sqrt{\Delta_3(P_i,p_a,P_{i+1})\cdot\Delta_3(P_i,p_a,P_{i-1})}}{\lambda(s_i,m_a^2,t_i)}\eqcomma
\label{eq:s_to_phi}
\end{equation}
where $\Delta_3$ denotes the Gram determinant of three momenta. Explicitly, for three momenta $(k_1,k_2,k_3)$, we obtain
\begin{equation}
\Delta_3(k_1,k_2,k_3)
=\frac{1}{8}
\begin{vmatrix}
2 k_1^2 & 2 k_1\!\cdot\! k_2 & 2 k_1\!\cdot\! k_3 \\[0.1cm]
2 k_2\!\cdot\! k_1 & 2 k_2^2 & 2 k_2\!\cdot\! k_3 \\[0.1cm]
2 k_3\!\cdot\! k_1 & 2 k_3\!\cdot\! k_2 & 2 k_3^2
\end{vmatrix}\eqperiod
\end{equation}
The quantity $V$ appearing in Eq.\eqref{eq:s_to_phi} is given by
\begin{equation}
V=-\frac{1}{8}
\begin{vmatrix}
2 s_i & s_i+m_a^2-t_i & s_i+s_{i-1}-m_i^2 \\[0.1cm]
m_a^2+ s_i-t_i & 2m_a^2 & m_a^2+s_{i-1}-t_{i-1} \\[0.1cm]
s_{i+1}+s_i-m_{i+1}^2 &   s_{i+1}+m_a^2-t_{i+1}   & 0          
\end{vmatrix}\eqperiod
\end{equation}
To cast Eq.\eqref{eq:byckling} into the standard unit-hypercube form, we define the mappings
\begin{equation}
\tilde{s}_i = \tilde{s}_i^{\min} + (\tilde{s}_i^{\max}-\tilde{s}_i^{\min})\,r_s
\qquad \mand \qquad
|t_{i-1}|  = G_{\nu}(r_t,0,-t_{i-1}^{\max},-t_{i-1}^{\min}) \eqcomma
\end{equation}
where the kinematic limits are determined by Eqs.\eqref{eq:s_to_phi} and~\eqref{eq:t_invariant_theta_23}.

%%===================================
\subsubsection{Three-particle decay phase space}
\label{sec:one_to_three}

We next consider a three-body decay $p_0=p_1+p_2+p_3$ with fixed parent mass $m_0^2=p_0^2$ and daughter masses $m_i$, as illustrated in Fig.~\ref{fig:diagrams_simple_blocks}.
In the rest frame of $p_0$, the phase space is five-dimensional and can be parametrized by two independent energies and three angles.
Unlike the two-particle decay, the energies of the final-state particles are no longer fixed by kinematics and must be treated as integration variables.
Following Ref.~\cite{Knippen:2019ojd}, we choose the energies $E_1$ and $E_2$ of particles 1 and 2, together with the angles $(\theta,\phi)$ specifying the direction of $\vec p_1$ and an additional azimuthal angle $\beta$ describing the orientation of $\vec p_2$ around $\vec p_1$. The spatial parts of the first two momenta can be written as
\begin{equation}
\begin{alignedat}{2}
    \vec{p}_1&= |\vec{p}_1|\, \mathcal{R}(\phi,\theta)\,(0,0,1)^\mathrm{T} & \qquad \mwith \qqquad |\vec{p}_1|&=\sqrt{E_1^2-m_1^2}\eqcomma\\
    \vec{p}_2&= |\vec{p}_2| \,\mathcal{R}(\phi,\theta)\,\mathcal{R}(\beta,\alpha)\,(0,0,1)^\mathrm{T} & \qquad |\vec{p}_2|&=\sqrt{E_2^2-m_2^2}\eqcomma
\end{alignedat}
\end{equation}
where $\mathcal{R}(\phi,\theta)$ denotes the rotation matrix
\begin{equation}
\mathcal{R}(\phi,\theta)
=
\begin{pmatrix}
\cos\phi & -\sin\phi & 0\\
\sin\phi &  \cos\phi & 0\\
0        &  0        & 1
\end{pmatrix}
\begin{pmatrix}
 \cos\theta & 0 & \sin\theta\\
 0          & 1 & 0\\
 -\sin\theta& 0 & \cos\theta
\end{pmatrix}\eqperiod
\end{equation}
The polar angle $\alpha$ between $\vec p_1$ and $\vec p_2$ is fixed by energy–momentum conservation and is given by
\begin{equation}
\cos\alpha
= \frac{2m_0\!\left(\frac{m_0}{2}-E_1-E_2\right)
      + m_1^2+m_2^2+2E_1E_2-m_3^2}
       {2|\vec p_1||\vec p_2|}\eqperiod
\end{equation}
The allowed energy ranges are
\begin{equation}
\begin{split}
E_1^\mathrm{max}&=\frac{m_0}{2}+\frac{m_1^2-(m_2+m_3)^2}{2m_0}\\
E_2^\mathrm{min/max}&=\frac{1}{2\Delta}\left[(m_0-E_1)(\Delta+\Delta_{23})\mp\sqrt{|\vec{p}_1|^2\left((\Delta+\Delta_{23})^2-4m_2^2\Delta\right)}\right]\eqcomma
\end{split}
\end{equation}
with the shorthand
\begin{equation}
    \Delta=2m_0\left(\frac{m_0}{2}-E_1\right)+m_1^2\eqperiod
\end{equation}
We can then introduce the mappings
\begin{equation}
\begin{alignedat}{2}
E_1 &= m_1 + (E^{\max}_1-m_1)\,r_{E_1}\eqcomma & E_2 &= E^{\min}_2 + (E^{\max}_2-E^{\min}_2)\,r_{E_2}\eqcomma \\
\phi &= 2\pi r_\phi\eqcomma \qqquad
\beta = 2\pi r_\beta\eqcomma &\qqquad
\cos\theta &= 2r_\theta-1 \eqperiod
\end{alignedat}
\end{equation}
The momenta $p_{i}$ are constructed from these variables in the decay rest frame and boosted back to the frame defined by $p_0$. The phase-space measure can be written in terms of the physical variables as
\begin{equation}
\begin{split}
\int \d\Phi_3(x)
&=
\frac{1}{8}
\int_{m_1}^{E_1^{\max}} \d E_1
\int_{E_2^{\min}}^{E_2^{\max}} \d E_2
\int_0^{2\pi} \d\phi
\int_0^{2\pi} \d\beta
\int_{-1}^{1} \d\!\cos\theta \eqcomma
\end{split}
\end{equation}

%%===================================
\subsubsection{$s$-channel integration order}
\label{sec:s_order}

To combine the elementary $1\to2$ and $1\to3$ decays into the full $s$-channel part of the phase-space mapping, it is necessary to specify the corresponding integration boundaries. These boundaries depend on the order in which the
invariant masses $s_i$ are sampled.

We consider the set of invariants $s_i$ associated with a given decay tree, excluding those included in the $t$-channel parametrization. The root of the tree is given by the squared partonic center-of-mass energy $\hat{s}$, while
the leaves correspond to the outgoing on-shell particle masses. If present, the $t$-channel contribution is treated as a single effective node with $\kappa + 1$ children.

The index $i$ labels the sequence in which the invariants are generated. For a given ordering, the allowed range of each $s_i$ is determined by the kinematics and by the values of previously generated invariants. This naturally leads to
the following recursive procedure:
\begin{enumerate}
    \item Set $i=1$.

    \item Recursively determine the minimal masses for all nodes $j$ of the decay tree, starting from the leaves. For an outgoing particle with mass $m_j$, set $m_{\text{min},j} = m_j$. For a node with an invariant $s_j$ that has already been sampled, set $m_{\text{min},j} = \sqrt{s_j}$. Otherwise, the minimal mass is given by the sum of the minimal masses of its children,
    \begin{equation}
        m_{\text{min},j}
        = \sum_{k\,\in\,\text{child of } j} m_{\text{min},k} \eqperiod
    \end{equation}

    \item Starting from node $i$, move towards the root of the tree until a node with an already sampled invariant $s_r$ is found. Along this path, collect the set $\cal{J}$ of all nodes that branch off from the visited nodes, including the children of node $r$, but excluding the children of node $i$. The maximal allowed mass is then given by
    \begin{equation}
        m_{\text{max},i}
        = \sqrt{s_r}
        - \sum_{j \in \mathcal{J}} m_{\text{min},j} \eqperiod
    \end{equation}

    \item Sample the invariant $s_i$ within the integration boundaries
    \begin{equation}
        s_{\text{min},i} = m_{\text{min},i}^2
        \qquad \mand \qquad
        s_{\text{max},i} = m_{\text{max},i}^2 \eqperiod
    \end{equation}
    \item Increase $i \to i + 1$ and return to step~2 until all invariants have been sampled.
\end{enumerate}
The freedom in choosing the integration order makes it possible to handle even non-trivial decay topologies, such as
$\PH \to \PZ\PZ \to \Pq \Pqbar \Pq \Pqbar$, where at most two of the three propagators can be on shell at the same time and the choice of integration order therefore has a significant impact on the resulting phase-space distribution. In a multi-channel setup we account for such cases by building multiple sub-channels for all possible on-shell configurations, where the propagators that can be on shell are sampled first. These sub-channels are then weighted by the denominators of their respective on-shell propagators.

%%===================================
\subsubsection{$t$-channel integration order}
\label{sec:t_order}

As for the $s$-channel part of the mapping, there is some freedom in the choice of integration order for the $t$-channel invariants. For simplicity, and following Ref.~\cite{Mattelaer:2021xdr}, we restrict ourselves to integration orders in which the invariants are generated inwards, starting from the incoming legs of the diagram. 
This restriction reduces the number of possible
orders from $\kappa!$ to $2^{\kappa-1}$.

We first sample $\kappa-1$ time-like invariants $s_1, \ldots, s_{\kappa-1}$. We denote the two incoming parton momenta of the underlying $2\to n$ scattering process by $p_a$ and $p_b$, and the outgoing legs generated by the $t$-channel mapping by $p_1,\ldots,p_{\kappa+1}$ with corresponding masses $m_1,\ldots,m_{\kappa+1}$.
These masses can either correspond to on-shell final-state particles or to intermediate off-shell masses associated with previously sampled time-like invariants from $s$-channel decays.
The permutation $\sigma(i)$ specifies the order in which the outgoing momenta are generated.
For $\kappa > 1$, the invariants $s_i$ are determined using the following procedure:
\begin{enumerate}
    \item Set $i = 1$, and define $s_0 = \hat{s}$.
    
    \item Sample the invariant $s_i$ within the kinematic boundaries
    \begin{equation}
        s_{\text{min},i}
        = \left( \sum_{j=i+1}^{\kappa + 1} m_{\sigma(j)} \right)^2
        \qquad \text{and} \qquad
        s_{\text{max},i}
        = s_{i-1} - m_{\sigma(i)}^2 \eqperiod
    \end{equation}
    \item Increase $i \to i+1$ and return to step~2 until $i \leq \kappa-1$.
\end{enumerate}
Once the time-like invariants have been fixed, the space-like invariants and momenta are generated as follows:
\begin{enumerate}
    \item Set $i = 1$, $p_A = p_a$ and $p_B = p_b$.
    
    \item For $i < \kappa$, determine whether the particle $\sigma(i)$ is closer to incoming leg~$A$ or~$B$ than the remaining, not yet generated outgoing particles.
    
    \item Define the outgoing masses $\tilde{m}_1$ and $\tilde{m}_2$ of the next $2\to 2$ scattering block according to
    \begin{equation}
    \begin{alignedat}{3}
        \tilde{m}_1 &= m_{\sigma(\kappa)} &\quad\mand\quad \tilde{m}_2 &= m_{\sigma(\kappa + 1)} \qquad&&\text{if } i=\kappa \\
        \tilde{m}_1 &= m_{\sigma(i)} &\quad\mand\quad \tilde{m}_2 &= \sqrt{s_i} \qquad&&\text{if closer to leg A} \\
        \tilde{m}_1 &= \sqrt{s_i} &\quad\mand\quad \tilde{m}_2 &= m_{\sigma(i)} \qquad&&\text{if closer to leg B} \eqperiod
    \end{alignedat}
    \end{equation}
    
    \item Sample the momenta $\tilde{p}_1$ and $\tilde{p}_2$ given $p_A$, $p_B$, $\tilde{m}_1$, and $\tilde{m}_2$, as
    described in Sec.~\ref{sec:two_to_two}.
    
    \item Assign the generated momenta and update the incoming legs,
    \begin{equation}
    \begin{alignedat}{2}
        p_{\sigma(i)} &= \tilde{p}_1 \quad\mand\quad p_{\sigma(i+1)} = \tilde{p}_2 \qquad&&\text{if } i=\kappa \\
        p_{\sigma(i)} &= \tilde{p}_1 \quad\mand\quad p_B \to p_B - \tilde{p}_2 \qquad&&\text{if closer to leg A} \\
        p_{\sigma(i)} &= \tilde{p}_2 \quad\mand\quad p_A \to p_A - \tilde{p}_1 \qquad&&\text{if closer to leg B} \eqperiod
    \end{alignedat}
    \end{equation}
    
    \item Increase $i \to i+1$ and return to step~2 until $i \leq \kappa$.
\end{enumerate}
If no explicit integration order is specified, a heuristic choice is made by prioritizing the most singular propagators, as estimated from the masses of the outgoing particles, following Ref.~\cite{Mattelaer:2021xdr}.

%%%%%%%%%%%%%%%%%%%%%%%%%%%%%%%%%%%%%%%%%%%%%%%%%%%
\subsection{Rambo and FastRambo}
\label{sec:fastrambo}

The \rambodiet~\cite{RamboDiet} and \hicom~\cite{vanHameren:2003pr} algorithms are invertible variants of the classic \rambo algorithm~\cite{Rambo}, both serving as simple phase-space generators that yield constant weights for massless final states. While the invertibility of \rambodiet is a valuable feature, particularly in modern phase-space generation pipelines, it requires numerically solving a polynomial equation. This is computationally impractical, especially when we aim to generate multiple events simultaneously in batches.

In many realistic applications, however, a perfectly flat phase space is usually not needed. This observation allows one to relax the exact flatness condition in favor of analytic simplicity and computational efficiency. The \hicom algorithm already follows this philosophy: Besided the flat variant, it also provides a fully analytic and invertible mapping that yields an approximately flat phase space, thereby avoiding the costly numerical inversion step of \rambodiet while remaining efficient to evaluate.
Following the same guiding principle, we replace the non-analytic polynomial inversion in \rambodiet with a fully analytical, invertible rational–quadratic transformation. Building on the original \rambodiet, we introduce the \fastrambo algorithm, which retains strict invertibility while eliminating the expensive numerical root-finding step. To motivate and explain this construction, we first review the standard \rambodiet derivation before outlining the minimal modifications required to obtain our analytical alternative.

We start by considering the $n$-body phase space for a total momentum $Q$ and omit again the overall factor of $(2\pi)^{4-3n}$ for readability
\begin{equation}
    \d\Phi_n\left(\{p_a,m_1\},\dots,\{p_n,m_n\} \,\big|\, Q\right)
    = \delta^{(4)}\!\left( \sum_{i=1}^n p_i - Q \right) 
       \prod_{i=1}^n \d^4 p_i \,
       \delta\!\left(p_i^2 - m^2_i\right) \,
       \Theta(p_i^0 - m_i) \eqperiod
    \label{eq:phase_space_measure}
\end{equation}
This can be factorized iteratively into $1 \to 2$ decays as
\begin{equation}
\begin{split}
\d\Phi_n\left(\{p_a,m_1\},\dots,\{p_n,m_n\} \,\big|\, Q\right)
&= \left( \prod_{i=2}^n 
    \d\Phi_2\big(\{p_{i-1},m_{i-1}\}, \{Q_i,M_i\} \,\big|\, Q_{i-1}\big) \right)\\
&\times \left( \prod_{i=2}^{n-1} 
    \Theta(M_{i-1} - m_{i-1} - M_i) \;
    \Theta\!\left(M_i-\sum_{k=i}^n m_k\right) \,
    \d M_i^2 \right) \eqcomma
\end{split}
\label{eq:full_factorization}
\end{equation}
where $M_i$ corresponds to the intermediate mass of a virtual particle. We further identify $Q_1=Q$, $M_1=\sqrt{Q^2}$, and $\{Q_n, M_n\}=\{p_n, m_n\}$.  We can parametrize the two–body decay in the rest frame of $Q_{i-1}$ as
\begin{equation}
  \d\Phi_2\!\left(\{p_{i-1},m_{i-1}\},\{Q_i,M_i\}| Q_{i-1}\right)
  = \rho(M_{i-1},M_i,m_{i-1})\;\d\!\cos\theta_{i-1}\,\d\phi_{i-1}\,,
  \label{eq:dphi2}
\end{equation}
with the two–body density factor
\begin{equation}
  \rho(M_{i-1},M_i,m_{i-1})
  = \frac{1}{8M_{i-1}^2}\sqrt{\lambda\!\left(M_{i-1}^2,M_i^2,m^2_{i-1}\right)}\eqcomma
\end{equation}
We can then absorb the two-body density factors and the intermediate masses into a mass measure given by
\begin{equation}
\begin{split}
\d M_n(M_2,\dots,M_{n-1}|M_1;m_1,\dots,m_n)&=
\rho(M_{n-1},M_n,m_{n-1})\\
\times
\left( \prod_{i=2}^{n-1} \rho(M_{i-1},M_i,m_{i-1})
        \right.&\left.\Theta(M_{i-1} - m_{i-1} - M_i) \;
        \Theta\!\left(M_i-\sum_{k=i}^n m_k\right)\,
        \d M_i^2 \right)\eqcomma
\end{split}
\label{eq:mass_measure}
\end{equation}
which is connected to the total $n$-body phase space by
\begin{equation}
\begin{split}
\d\Phi_n\left(\{p_a,m_1\},\dots,\{p_n,m_n\} \,\big|\, Q\right)
&= \d M_n(M_2,\dots,M_{n-1}|M_1;m_1,\dots,m_n)\\
&\quad\times \left(\prod_{i=2}^n \d\!\cos\theta_{i-1}\,\d\phi_{i-1}\right)\eqperiod
\end{split}
\label{eq:n_body_mass_factor}
\end{equation}
To simplify the presentation of the following formulas, we first consider the massless case. In this massless limit, Eq.\eqref{eq:mass_measure} simplifies to
\begin{equation}
\d M_n(M_2,\dots,M_{n-1}|M_1;0,\dots,0)=\frac{1}{8^{n-1}}\prod_{i=2}^{n-1} \frac{M^2_{i-1}-M^2_{i}}{M^2_{i-1}}\,\Theta(M^2_{i-1} - M^2_i) \;\Theta(M^2_i) \;\d M_i^2\eqperiod
\label{eq:mass_measure_massless}
\end{equation}
Now, expressing $M_i=u_2\dots u_i\,M_1$ maps the integration variables onto the unit-hypercube $u_i\in[0,1]$ and allows us to write
\begin{equation}
\d M_n(M_2,\dots,M_{n-1}|M_1;0,\dots,0)=\frac{M_1^{2n-4}}{8^{n-1}}\prod_{i=2}^{n-1} u_i^{2(n-i-1)}\,(1-u_i^2)\;\d u^2_i\eqperiod
\label{eq:mass_measure_u} 
\end{equation}
We can further simplify this equation by substituting
\begin{equation}
    y_i = u_i^{2(n-i)} \qquad \mwith \qquad
    \d y_i = (n-i)\,u_i^{2(n-i-1)}\,\d u_i^2\eqcomma
\end{equation}
which simplifies Eq.\eqref{eq:mass_measure_u} to
\begin{equation}
\d M_n(M_2,\dots,M_{n-1}|M_1;0,\dots,0)=\frac{M_1^{2n-4}}{8^{n-1}}\frac{1}{\Gamma(n-1)}\prod_{i=2}^{n-1}\,(1-y^{\frac{1}{n-i}}_i)\;\d y_i\eqcomma
\label{eq:mass_measure_x} 
\end{equation}
where $\Gamma(\cdot)$ denotes the gamma function.
Now, in the standard \rambodiet approach, these $y_i$ are then mapped onto random numbers $r_i$ by the mapping
\begin{equation}
    r_i \equiv G_\text{diet}(y_i, k)= (k+1)\,y_i - k\,y_i^{1+1/k}\qquad \mwith \qquad k=n-i\eqperiod
    \label{eq:r_to_x_orig}
\end{equation}
which leads to the known flat result
\begin{equation}
    \d M_n(M_2,\dots,M_{n-1}|M_1;0,\dots,0)=\frac{M_1^{2n-4}}{8^{n-1}}\frac{1}{\Gamma(n)\,\Gamma(n-1)}\prod_{i=2}^{n-1}\;\d r_i\eqperiod
    \label{eq:mass_measure_flat} 
\end{equation}
In general, to obtain $y_i$ from $r_i$ during sampling, Eq.\eqref{eq:r_to_x_orig} must be inverted. For $n \le 5$ this inversion, in principle, has an analytic solution, which is only straightforward for $n \le 3$ (quadratic case). 
For larger $n$ the formulas become cumbersome, numerically unstable, and require extra logic to select the physical root. Thus, numerical methods have to be used in most cases.\medskip

%------------------------------------------------------
\begin{figure}[t!]
    \centering
    \includegraphics[width=0.48\linewidth]{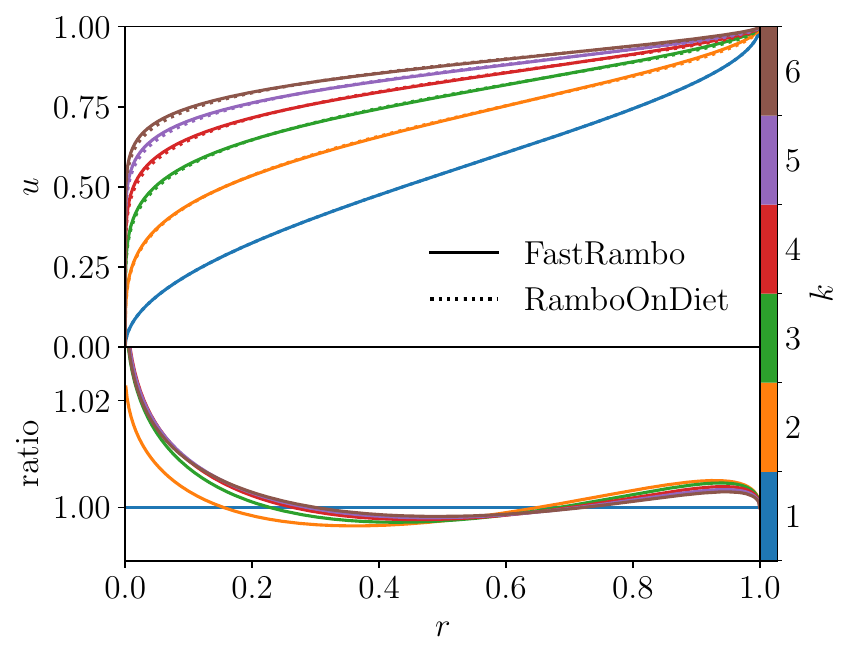}
    \includegraphics[width=0.495\linewidth]{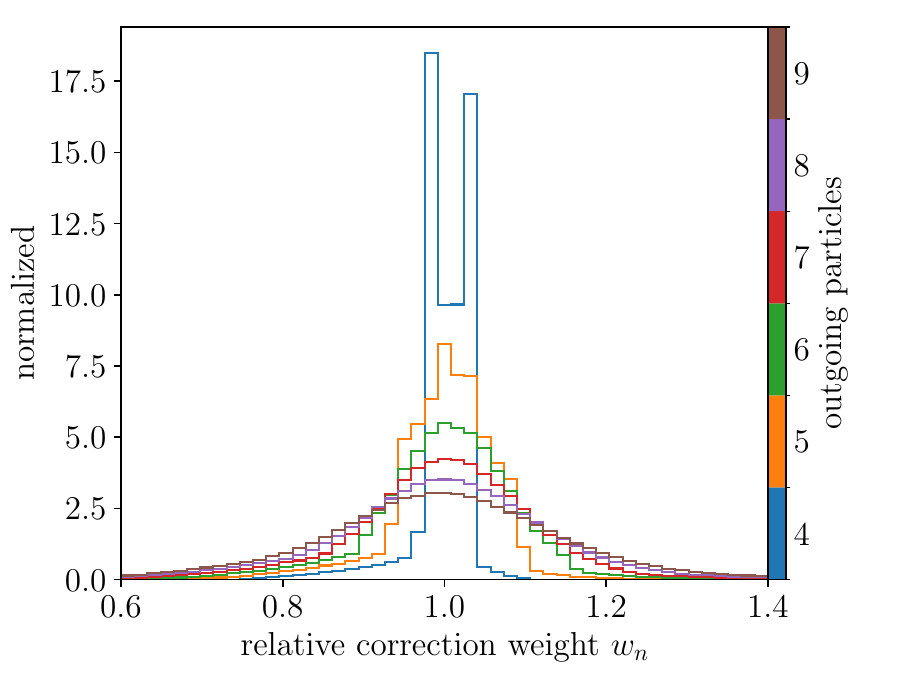}
    \caption{Left: Inverse RQF mapping compared to the original \rambodiet mapping. Right: distribution of \fastrambo phase-space weights normalized by the phase-space volume for a fixed center-of-mass energy of 1 TeV.}
    \label{fig:rambo_weights}
\end{figure}
%------------------------------------------------------

If uniform weights are not strictly required, the mapping in Eq.\eqref{eq:r_to_x_orig} can be
replaced by a rational quadratic function (RQF). Even a one-parameter RQF with a tunable endpoint slope is sufficiently expressive and has the advantage of having a closed-form inverse and Jacobian. We parametrize the RQF by a positive parameter $c_k>0$, defined as the derivative 
at the lower ($y_i=0$) endpoint of the mapping. 
As the derivative at $y_i=1$ is zero for all \rambodiet mappings, we also fix this derivative for our new RQF transformation. The new mapping is thus given by~\cite{durkan2019neural}
\begin{equation}
\begin{split}
r'_i \equiv G_\text{RQF}(y_i, c_k) &= \frac{y_i^2 + c_k\,y_i(1-y_i)}
             {1 + (c_k-2)\,y_i(1-y_i)} \\
g_\text{RQF}(y_i, c_k) &= \frac{2y_i\,(1-y_i)+c_k\,(1-y_i)^2}
{\left[\,1 + (c_k-2)\,y_i\,(1-y_i)\,\right]^2} \eqcomma
\end{split}
\end{equation}
where the parameter $c_k$ should be chosen depending on the value of $k$ in Eq.\eqref{eq:r_to_x_orig}. The mapping is strictly monotonic on $[0,1]$, and has a closed-form inverse which follows from solving a quadratic equation~\cite{durkan2019neural}. 
In our single-bin case, this can be written compactly as
\begin{equation}
y_i \equiv G_\text{RQF}^{-1}(r'_i, c_k) \;=\;
\frac{2r'_i}{b_k + \sqrt{b_k^{2} + 4r'_i\,(1-b_k)}} \qquad \mwith \qquad
b_k = c_k - (c_k - 2)\,r'_i \eqperiod
\label{eq:r_to_y_inverse}
\end{equation}
This expression is numerically stable for all $r'_i\in[0,1]$ and $c_k>0$, and avoids an expensive numerical inversion step.
To approximate the original \rambodiet mapping as close as possible, we determine the spline parameter for each $k=1,\dots,n-2$ by a one-time fit
\begin{equation}
c^\star_k=\arg\min_{c>0}\int_0^1 \d y\; \left[G_\text{RQF}(y, c)-G_\text{diet}(y, k)\right]^2\eqperiod
\end{equation}
This tuning is performed only \emph{once} per $k$ and the resulting values $c^\star_k$ are then simply stored in a small lookup table. The results for $k=1 \ldots 7$ are given in Tab.~\ref{tab:rambo_fit}. During phase-space sampling no further optimization is needed. This yields an analytic, easily invertible mapping that closely resembles the standard \rambodiet mapping. With the replacement $y_i = G_{c^\star}(r'_i)$ and $k=n-i$, the new mass measure is now given by
\begin{equation}
\begin{split}
\d M_n(M_2,\dots,M_{n-1}|M_1;0,\dots,0)
&=\frac{M_1^{2n-4}}{8^{n-1}}\frac{1}{\Gamma(n-1)}\prod_{i=2}^{n-1}\frac{(1-y_i^{\frac{1}{k}})}{g_\text{RQF}(y_i, c_{k})}\;\d r'_i \\
&\equiv
\frac{M_1^{2n-4}}{8^{n-1}}\frac{w_{n}(r)}{\Gamma(n)\,\Gamma(n-1)}\prod_{i=2}^{n-1}\;\d r'_i\eqperiod
\end{split}
\label{eq:mass_measure_rqs} 
\end{equation}
In this formulation the phase-space weights are no longer perfectly uniform and acquire an additional local factor $w_{n}(r)$. The resulting distribution of $w_{n}(r)$ for a fixed center-of-mass energy of 1000 GeV is shown in Fig.~\ref{fig:rambo_weights}. The deviation from flatness can, however, be minimized by an appropriate choice of $c_k^\star$. We note that for $k=1$ and $c^\star_{k=1}=2$, the RQF mapping coincides with the original \rambodiet mapping.

%-----------------------------------
\begin{table}[t!]
    \setlength{\tabcolsep}{10pt}
    \centering
    \begin{small} \begin{tabular}{l|ccccccc}
        \toprule
         $k$ & 1 & 2 & 3 & 4 & 5 & 6 & 7 \\
         \midrule
         $c_k^*$ & 2 & 2.712001 & 3.084521 & 3.313073 & 3.467512 & 3.578833 & 3.662872 \\
         \bottomrule
    \end{tabular} \end{small}
    \caption{Fitted parameters $c_k^*$ of the \fastrambo mapping for $k=1 \ldots 7$.}
    \label{tab:rambo_fit}
\end{table}
%------------------------------------

%%===================================
\subsubsection{Massive case}

The construction presented above assumes massless external particles. The extension to massive final states can be performed via two conceptually different strategies: (i) a rescaling of momenta obtained by solving a set of kinematic constraints to enforce the on-shell conditions for massive particles~\cite{Rambo}, and (ii) an alternative formulation in which mass effects are incorporated through a reweighting of massless phase-space points~\cite{RamboDiet}.

In \madspace, we adopt the reweighting approach~\cite{RamboDiet}, which avoids the need to solve the non-linear rescaling equations and is therefore particularly well suited for a compute-graph-based and fully vectorized realization. Starting from a massless phase-space point with momenta $p_i$, the corresponding massive configuration with masses $m_i$ is obtained by a global rescaling with a weight factor
\begin{equation}
w(p_i,m_i) 
= \frac{1}{8}\prod_{i=2}^n\frac{\rho(M_{i-1},M_i,m_{i-1})}{\rho(K_{i-1},K_i, 0)}\; \prod_{i=2}^{n-1}
\frac{M_i}{K_i}\eqcomma
\label{eq:rambo_mass_reweight}
\end{equation}
with
\begin{equation}
K_i = M_i-\sum_{j=i}^n m_j \qquad \mfor \qquad i=1,\dots,n-1 \qquad \mand \qquad K_n=0
\end{equation}
This factor accounts for the Jacobian relating the massive and massless phase-space measures and reproduces the correct Lorentz-invariant $n$-body phase space in the presence of non-vanishing particle masses~\cite{Rambo,RamboDiet}.

%%%%%%%%%%%%%%%%%%%%%%%%%%%%%%%%%%%%%%%%%%%%%%%%%%%
\subsection{Chili}
\label{sec:chili_mapping}

In addition to the mappings discussed above, \madspace also includes the \chili~\cite{Bothmann:2023siu,Bothmann:2023gew} mapping. 
Conceptually, \chili is closely related to the phase-space construction employed in \alpgen~\cite{Mangano:2002ea}, where multi-parton final states are generated directly in terms of transverse momenta, rapidities, and azimuthal angles, with the incoming momentum fractions reconstructed from the final-state kinematics.
Using collider coordinates, and dropping overall factors of $(2\pi)^{4-3n}$, we can write the $n$-body phase space as
\begin{equation}
\d\Phi_n(x) = \delta^{(4)}\!\left( p_a+p_b - \sum_{i=1}^n p_i \right)\left[\prod_{i=1}^n\frac{\d p^2_{\text{T},i}\,\d y_i\,\d\phi_i}{4}\right]
\eqperiod
\label{eq:chili_one_particle_measure}
\end{equation}
The combination with PDFs and the analytic treatment of the energy-momentum conservation, lead to a construction in which the first $n-1$ final-state momenta are generated explicitly~\cite{Bothmann:2023siu}.
The transverse components of the last momentum are then fixed by recoil, while its remaining longitudinal degree of freedom is sampled in terms of its rapidity.

%\subsubsection{Sampling of first $n-1$ particles}

For each of the first $n-1$ particles, we first sample the transverse momentum using two possible mappings.
We set the maximal transverse momentum to
\begin{equation}
    p_{\text{T},\max} = \frac{\sqrt{s_\text{lab}}}{2} \; ,
\end{equation}
unless it is limited by an upper cut.
If a lower cut $p_\text{T}>p_{\text{T},\min}$ is applied, we sample $p_\text{T}^2$ from an inverse distribution,
\begin{equation}
\begin{split}
p_\text{T}^2 =
\left(\frac{r}{p_{\text{T},\max}^2}
      +\frac{1-r}{p_{\text{T},\min}^2}\right)^{-1}
\qquad \mwith \qquad g_\text{cut}(p_\text{T}^2,p_{\text{T},\min},p_{\text{T},\max})
=\frac{p_\text{T}^2}{p_{\text{T},\min}^2}-\frac{p_\text{T}^2}{p_{\text{T},\max}^2}\eqperiod  
\end{split}
\label{eq:chili_pt_withcut}
\end{equation}
which corresponds to a $1/p_\text{T}^2$-type sampling between $p_{\text{T},\min}^2$ and $p_{\text{T},\max}^2$. 
If no minimum $p_\text{T}$ cut is applied, we use a mapping that regulates the small-$p_\text{T}$ region by some energy scale $\lambda_C$,
\begin{equation}
p_\text{T}
=
\frac{2 \lambda_C\,p_{\text{T},\max}\,r}{2\lambda_C+p_{\text{T},\max}(1-r)}
\qquad \mwith \qquad g_\text{no-cut}(p_{\text{T}},\lambda_C,p_{\text{T},\max})
=
\frac{p_{\text{T}}\,p_{\text{T},\max}\,\bigl(2\lambda_C+p_\text{T}\bigr)^2}{2\lambda_C^2+\lambda_C\,p_{\text{T},\max}}\eqcomma
\label{eq:chili_pt_nocut}
\end{equation}
For massive particles, we simply choose $\lambda_C=m$, and for massless particles we set $m=1$ as default. The maximum for the absolute value of the rapidity as a function of the sampled transverse momentum is given by
\begin{equation}
    y_{\max} = \log\left(
      \sqrt{\frac{s_\text{lab}}{4 m_\text{T}^2}}
      + \sqrt{\frac{s_\text{lab}}{4 m_\text{T}^2} - 1}
    \right) \qquad \mwith \qquad
    m_\text{T}^2 = p_\text{T}^2 + m^2\;.
\end{equation}
This range can be further restricted if a rapidity cut is specified. We sample the rapidity and azimuthal angle as
\begin{equation}
y = y_{\max}\,(2r_{y}-1)\eqcomma
\qquad
\phi = 2\pi r_{\phi} + \phi_\text{rec}\eqcomma
\label{eq:chili_phi_relative}
\end{equation}
where the azimuthal angle is sampled relative to the current recoil direction $\phi_\text{rec}$ at each step, which reduces degeneracies in the azimuthal orientation and improves numerical stability for large multiplicities.
After generating the first $n-1$ momenta, we fix
\begin{equation}
p_{x,n}=-\sum_{i=1}^{n-1} p_{x,i}  \qquad \mand \qquad p_{y,n}=-\sum_{i=1}^{n-1} p_{y,i}  \eqperiod
\end{equation}
and sample the rapidity $y_n$ uniformly within the kinematic range implied by the already generated subsystem. The initial-state momenta can then simply be constructed from the final-state momenta as
\begin{equation}
E^\pm = \sum_{i=1}^n E_{i}\pm \sum_{i=1}^n p_{z,i}
\qqquad
p_a =\left(\frac{E^+}{2},0,0,\frac{E^+}{2}\right)
\qqquad
p_b=\left(\frac{E^{-}}{2},0,0,-\frac{E^{-}}{2}\right)\eqperiod
\end{equation}
Note that the \chili mapping does not guarantee physical momentum configurations as the resulting $E^+$ and $E^-$ can exceed the beam energy. To obtain the correct physical results, these phase-space points have to be removed using a technical cut.

\clearpage
%%%%%%%%%%%%%%%%%%%%%%%%%%%%%%%%%%%%%%%%%%%%%%%%%%%
\section{The MadSpace framework}
\label{sec:framework}

\madspace is a stand-alone library for parton-level phase-space integration and event generation, with a particular focus on scalability, hardware portability, and extensibility towards modern sampling and inference techniques.
While it currently supports only LO computations, its architecture is designed to accommodate a much broader set of applications in the future.
Rather than optimizing for a single use case or process class, the framework aims to provide a unified execution model that can accommodate traditional Monte Carlo integration, adaptive importance sampling, and fully ML-driven approaches within the same infrastructure.

Within \texttt{MadGraph}, \madspace serves as the phase-space and event-generation layer.
Model handling based on the Universal FeynRules Output (UFO)~\cite{Degrande:2011ua,Darme:2023jdn}, process and diagram generation, and the generation of matrix-element code remain the task of \texttt{MadGraph} and are not modified by \madspace.
\madspace only receives the information required to construct the phase space, namely the Feynman-diagram topologies and the masses and widths of all external and internal particles, and it evaluates the matrix elements through the \umami interface described in Sec.~\ref{sec:umami}.
Consequently, any process in any UFO model for which \texttt{MadGraph} generates \umami-compatible matrix elements can be integrated with \madspace without further modification.
Outside of \texttt{MadGraph}, \madspace can be combined with any matrix-element implementation that provides the \umami interface, and its individual building blocks can be used directly through the \python interface described in Sec.~\ref{sec:python}.

At the core of this design lies a strict separation between physics logic and numerical execution.
All physics-specific aspects, such as phase-space decomposition, channel definitions, and observable construction, are encoded at graph construction time, while the actual event generation is performed by executing a precompiled compute graph on batches of events.

%%%%%%%%%%%%%%%%%%%%%%%%%%%%%%%%%%%%%%%
\subsection{Technical implementation}

Phase-space samplers in parton-level event generators must balance flexibility, algorithmic complexity, and runtime overhead. Modern generators such as \mg and \comix~\cite{Gleisberg:2008fv} in \sherpa achieve a high degree of flexibility by exploiting the topology of the underlying Feynman diagrams within a multi-channel approach.
In \mg, this is achieved by generating dedicated phase-space code compiled separately for each process, whereas \comix employs a fully recursive strategy for phase-space construction at runtime.
Other generators favor simpler and more limited mappings, such as \rambodiet in \herwig or \chili in \pepper, trading flexibility for reduced implementation complexity.

\madspace follows a different design philosophy based on a compute-graph execution model, inspired by deep learning frameworks such as \tensorflow (static graph execution) and \pytorch (TorchScript). 
This approach enables a strict separation between physics-specific logic and the hardware-dependent numerical implementation. 
All phase-space mappings, sampling strategies, and auxiliary operations are expressed as a directed acyclic graph of elementary operations acting on tensors that represent batches of events.
Each operation consumes immutable input tensors and produces one or more output tensors, without modifying global program states.
This functional design naturally enables parallel execution and simplifies reasoning about correctness and reproducibility.
Complex tasks, such as the analysis of Feynman-diagram topologies and the assembly of multi-channel phase-space mappings, are performed only once during graph construction.
The resulting compute graph is stored in memory in an efficient byte-code representation and executed by a lightweight interpreter.
Since all operations are vectorized over batches of events, the runtime overhead of graph interpretation is negligible in typical integration and event-generation workloads.

%------------------------------------------------------
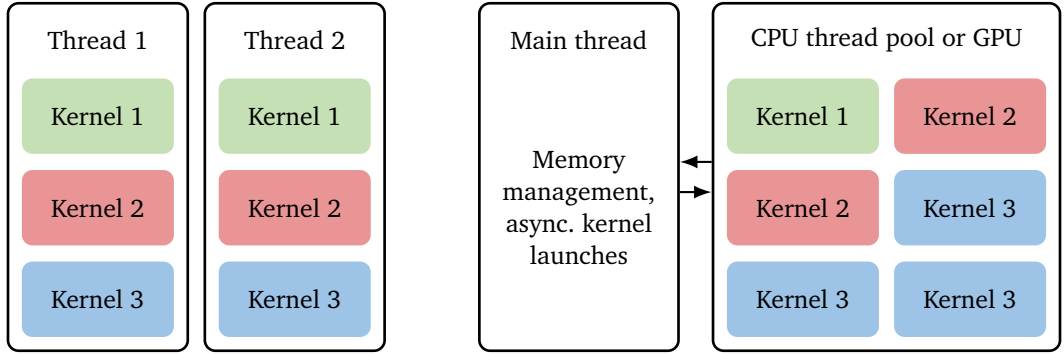
\begin{figure}[t!]
    \centering
    \begin{tikzpicture}[font=\small]

\node (sync_thread1) [minimum height=0.6cm] {Thread 1};
\node (sync_job11) [rectangle, rounded corners, fill=Gcolor, minimum width=2cm, minimum height=1cm, align=center, below=0.2cm of sync_thread1] {Kernel 1};
\node (sync_job12) [rectangle, rounded corners, fill=Rcolor, minimum width=2cm, minimum height=1cm, align=center, below=0.2cm of sync_job11] {Kernel 2};
\node (sync_job13) [rectangle, rounded corners, fill=Bcolor, minimum width=2cm, minimum height=1cm, align=center, below=0.2cm of sync_job12] {Kernel 3};
\node (sync_box1) [
  draw,
  rounded corners,
  line width=1pt,
  fill=none,
  inner sep=5pt,
  fit=(sync_thread1) (sync_job11) (sync_job12) (sync_job13)] {};

\node (sync_thread2) [minimum height=0.6cm, right=1cm of sync_thread1] {Thread 2};
\node (sync_job21) [rectangle, rounded corners, fill=Gcolor, minimum width=2cm, minimum height=1cm, align=center, below=0.2cm of sync_thread2] {Kernel 1};
\node (sync_job22) [rectangle, rounded corners, fill=Rcolor, minimum width=2cm, minimum height=1cm, align=center, below=0.2cm of sync_job21] {Kernel 2};
\node (sync_job23) [rectangle, rounded corners, fill=Bcolor, minimum width=2cm, minimum height=1cm, align=center, below=0.2cm of sync_job22] {Kernel 3};
\node (sync_box2) [
  draw,
  rounded corners,
  line width=1pt,
  fill=none,
  inner sep=5pt,
  fit=(sync_thread2) (sync_job21) (sync_job22) (sync_job23)] {};
  
\end{tikzpicture}%
\hspace{3em}
\begin{tikzpicture}[font=\small]

\node (async_main) [minimum height=0.6cm] {Main thread};
\node (async_control) [minimum height=3.428cm, below=0.2cm of async_main, align=center] {Memory\\management,\\async.\ kernel\\launches};

\node (async_box1) [
  draw,
  rounded corners,
  line width=1pt,
  fill=none,
  inner sep=5pt,
  fit=(async_main) (async_control)] {};

\node (async_pool) [minimum height=0.6cm, right=1.1cm of async_main] {CPU thread pool or GPU};
\node (async_job11) [rectangle, rounded corners, fill=Gcolor, minimum width=2cm, minimum height=1cm, align=center, below=0.2cm of async_pool, xshift=-1.1cm] {Kernel 1};
\node (async_job12) [rectangle, rounded corners, fill=Rcolor, minimum width=2cm, minimum height=1cm, align=center, below=0.2cm of async_job11] {Kernel 2};
\node (async_job13) [rectangle, rounded corners, fill=Bcolor, minimum width=2cm, minimum height=1cm, align=center, below=0.2cm of async_job12] {Kernel 3};
\node (async_job21) [rectangle, rounded corners, fill=Rcolor, minimum width=2cm, minimum height=1cm, align=center, below=0.2cm of async_pool, xshift=1.1cm] {Kernel 2};
\node (async_job22) [rectangle, rounded corners, fill=Bcolor, minimum width=2cm, minimum height=1cm, align=center, below=0.2cm of async_job21] {Kernel 3};
\node (async_job23) [rectangle, rounded corners, fill=Bcolor, minimum width=2cm, minimum height=1cm, align=center, below=0.2cm of async_job22] {Kernel 3};
\node (async_box2) [
  draw,
  rounded corners,
  line width=1pt,
  fill=none,
  inner sep=5pt,
  fit=(async_pool) (async_job11) (async_job12) (async_job13) (async_job21) (async_job22) (async_job23)] {};

\draw [-Latex, thick] ([yshift=-0.2cm]async_box1.east) -- ([yshift=-0.2cm]async_box2.west);
\draw [-Latex, thick] ([yshift=0.2cm]async_box2.west) -- ([yshift=0.2cm]async_box1.east);
  
\end{tikzpicture}%
    \vspace{-2em}
    \caption{Illustration of the synchronous CPU graph execution mode (left) and asynchronous CPU and GPU execution modes (right).}
    \label{fig:exec_graph}
\end{figure}
%------------------------------------------------------

The concrete execution strategy depends on the target hardware.
On CPUs, \madspace supports two execution modes.
In asynchronous mode, a main thread interprets the compute graph and manages memory allocation, while individual operations are parallelized by submitting tasks to a thread pool. 
The number of tasks in the thread pool adapts dynamically to the batch size, making this mode particularly well-suited for scenarios in which multiple integration channels with varying event counts are evaluated concurrently, as in multi-channel \madnis training.
The fine-grained parallelism at the level of individual operations, however, comes with a higher runtime overhead.
As an alternative, we provide a synchronous execution mode that runs the complete graph operation sequence in a single thread.
Parallelization is then achieved by running multiple independent graph executions in parallel threads.
Since no synchronization is required between individual operations, this mode exhibits a lower runtime overhead at the expense of reduced flexibility.
We primarily use the synchronous mode for \vegas grid optimization and during event generation.
For GPUs, we implement a single execution mode that uses the asynchronous APIs of CUDA or HIP for most operations.
We illustrate the different execution modes in Fig.~\ref{fig:exec_graph}.
Most CPU-bound work can therefore be performed while the GPU is busy, resulting in very low runtime overhead.

%%%%%%%%%%%%%%%%%%%%%%%%%%%%%%%%%%%%%%%
\subsection{Features}

Our compute-graph-based approach is not limited to phase-space sampling. It is applied to the entire LO matrix-element-level event-generation chain, comprising random number generation, adaptive mappings such as \vegas or \madnis, phase-space mappings, cuts, multi-channel weight computation, PDF interpolation, matrix-element evaluation, and event unweighting.

%%================================
\subsubsection{Phase space}

Almost all phase-space mappings in \madspace are based on the topology of a tree-level Feynman diagram, including the graph structure, the masses of the incoming and outgoing particles, and the masses and widths of the propagators. Alternatively, we also provide the \fastrambo and \chili mappings, which do not encode a specific topology but can still be combined with $s$-channel decays to map out resonances explicitly.

\begin{figure}[t!]
    \centering
    \begin{minipage}[c]{0.30\linewidth}
        \centering
        \includegraphics[width=0.95\linewidth]{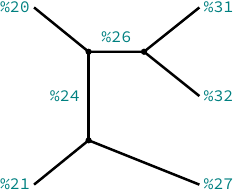}
    \end{minipage}
    \begin{minipage}[c]{0.68\linewidth}
    \centering
    \begin{lstlisting}[language=madgraph]
  Inputs:
    %0 : float[batch_size, 7]
  Instructions:
    %1, %2, %3, %4, %5, %6, %7 = unstack(%0)
    %11, %12 = stable_invariant_nu(%1, 0, 0.8, 0, 1.69e+08)
    %13 = sqrt(%11)
    %14, %15 = stable_invariant_nu(%2, 0, 0.8, 0, %11)
    %16 = sqrt(%14)
    %17, %18, %19 = r_to_x1x2(%3, %11, 1.69e+08)
    %20, %21 = com_p_in(%13)
    %22, %23 = t_inv_min_max(%20, %21, %16, 0)
    %24, %25 = stable_invariant_nu(%5, 0, 0.8, %22, %23)
    %26, %27, %28 = two_to_two_particle_scattering_com(
                    %4, %20, %21, %24, %16, 0)
    %29 = mul(%25, %28)
    %31, %32, %33 = two_body_decay(%6, %7, %16, 0, 0, %26)
    %34 = stack(%20, %21, %27, %31, %32)
    %35 = boost_beam(%34, %17, %18)
    %38 = stack(0.000102118, %12, %15, %19, %29, %33, 1)
    %39 = reduce_product(%38)
    %40 = cut_unphysical(%39, %35, %17, %18)
  Outputs:
    %35 : float[batch_size, 5, 4]
    %17 : float[batch_size]
    %18 : float[batch_size]
    %40 : float[batch_size]
    \end{lstlisting}
    \end{minipage}
    \caption{Compute graph generated by \madspace for an example diagram with both $t$-channel ({\fontfamily{SourceCodePro-TLF}\selectfont\color{teal}\%24}) and $s$-channel ({\fontfamily{SourceCodePro-TLF}\selectfont\color{teal}\%26}) propagators.}
    \label{fig:compute_graph}
\end{figure}

In Fig.~\ref{fig:compute_graph}, we illustrate the generated compute graph for an exemplary phase-space mapping of a simple hadronic $2 \to 3$ scattering process with one massless $s$-channel and one massless $t$-channel propagator. Numbers starting with a \verb|%| denote inputs, outputs, and intermediate variables. The individual operations directly correspond to the phase-space building blocks described in Sec.~\ref{sec:fd_mappings}.
The main components appearing in the example compute graph are:
\begin{itemize}
    \item \verb|stable_invariant_nu|: samples $\hat{s}$, $s$, and $|t|$ using Eq.\eqref{eq:prop_stable_nu},
    \item \verb|r_to_x1x2|: samples the momentum fractions of the incoming partons using Eq.\eqref{eq:pdf_sampling},
    \item \verb|t_inv_min_max| and \verb|two_to_two_particle_scattering_com|: generates the $2\to2$ scattering kinematics, see Sec.~\ref{sec:two_to_two},
    \item \verb|two_body_decay|: two-body decay of the intermediate resonance, see Sec.~\ref{sec:one_to_two},
    \item \verb|boost_beam|: boosts the generated momenta from the partonic center-of-mass frame to the laboratory frame.
\end{itemize}
%

%------------------------------------------------------
%------------------------------------------------------

%%================================
\subsubsection{Adaptive sampling}
\label{sec:adaptive}

To improve the weight distribution of generated phase-space points, \madspace provides built-in support for adaptive sampling techniques. This includes both the classic \vegas algorithm and neural importance sampling with \madnis. Within the compute-graph framework, adaptive samplers are implemented as interchangeable graph components that transform uniformly distributed random numbers into sampling distributions optimized for the target integrand.

For neural importance sampling, \madspace employs normalizing flows based on rational-quadratic spline transformations~\cite{durkan2019neural}.
The spline transformations are implemented via custom kernels that are fully integrated into the compute graph and evaluated in a vectorized manner, enabling efficient evaluation on both CPUs and GPUs.
The linear-algebra operations required by the neural networks are delegated to hardware-specific backend libraries, such as OpenBLAS on CPUs and CUBLAS or ROCBLAS on GPUs.
A detailed performance study of \madspace in combination with \madnis is deferred to a future publication.

%%================================
\subsubsection{PDF interpolation}
\label{sec:pdf}

To simplify installation and deployment, we design \madspace to be independent of any external libraries. To this end, we provide a built-in PDF interpolator that is compatible with the \lhapdf~\cite{Buckley:2014ana} grid format.
It can load PDF grids in the standard \lhapdf format, and its numerical output agrees with \lhapdf within 64 bit floating-point precision.

The interpolator is implemented for both CPUs and GPUs, enabling PDF evaluation on the same device as phase-space generation and matrix-element evaluation. This avoids unnecessary host–device data transfers and minimizes runtime overhead in heterogeneous CPU/GPU workflows.
The PDF interface is designed to be modular, so that alternative interpolation backends can be supported in the future without changing the surrounding event-generation logic.

%%================================
\subsubsection{Phase-space cuts}
\label{sec:cuts}

In \mg, cuts are defined using a fixed set of kinematic observables for fixed selections of particles. \madspace instead adopts a more flexible approach. In a first step, outgoing particles are selected by their PDG numbers. Multiple particle types can be combined, for instance to allow the selection of jets instead of gluons and individual quark flavours. The selection can comprise either all particles of a given type, or combinations such as pairs or triplets of particles of the same or different types (for example, pairs of jets or pairs of a jet and a lepton). Optionally, the selected momenta can be summed at this stage. In the second step, the observable of interest is computed. The currently available set of observables comprises:
\begin{itemize}
    \item \textbf{functions of a single momentum:} the four-momentum components $E$, $p_x$, $p_y$, $p_z$, the transverse momentum $p_\text{T}$, the magnitude of the three-momentum $|\vec{p}|$, the azimuthal angle $\phi$, the polar angle $\theta$ with respect to the beam axis, the rapidity and its absolute value $y$ and $|y|$, and the pseudorapidity and its absolute value $\eta$ and $|\eta|$;
    \item \textbf{functions of pairs of momenta:} the difference in azimuthal angle $\Delta\phi$, the difference in pseudorapidity $\Delta\eta$, the distance in the $(\eta,\phi)$ plane $\Delta R$, and the invariant mass $m$;
    \item \textbf{event-level observables:} the partonic center-of-mass energy $\sqrt{\hat{s}}$.
\end{itemize}
It is optionally possible to sum the computed observables over the selected particles. Finally, minimum and/or maximum values can be specified for each observable. If a cut is evaluated for multiple selected objects, it can be specified whether at least one or all of them are required to satisfy the cut.

%%================================
\subsubsection{Weighted histograms}
\label{sec:weighted_hists}

In addition to generating unweighted events, \madspace also supports creating weighted histograms during event generation. For a fixed number of integrand evaluations, such histograms exhibit smaller statistical uncertainties than those obtained from unweighted events, since the unweighting step discards part of the available weight information. Storing the full set of weighted events is typically prohibitive due to memory constraints. Instead, weighted histograms are therefore filled on the fly, with the binning fixed in advance.

The observables used for the weighted histograms in \madspace are defined using the same mechanism as for the phase-space cuts. They are evaluated on the same device as the phase-space generation and matrix-element computation. On GPUs, only the binned histogram data for each batch of events are transferred to host memory, thereby minimizing costly device–host data copies.

%%================================
\subsubsection{Renormalization and factorization scale choices}
\label{sec:scale}

We support the following choices for the renormalization and factorization scales:
\begin{itemize}
    \item fixed scale, often chosen as the sum of the final-state masses,
    \begin{equation}
        \mu_\text{R/F}=\sum_{i=1}^N m_i \; ;
    \end{equation}
    \item total transverse energy of the event,
    \begin{equation}
        \mu_\text{R/F}=E_\text{T}^\text{tot} = \sum_{i=1}^n \frac{E_i\,p_{\mathrm{T},i}}{|\vec{p}_i|} \; ;
    \end{equation}
    \item sum of the transverse masses (optionally divided by two),
    \begin{equation}
        \mu_\text{R/F}=H_\text{T}=\sum_{i=1}^n \sqrt{m_i^2 + p_{\mathrm{T},i}^2} \; ;
    \end{equation}
    \item partonic center-of-mass energy, \ie $\mu_\text{R/F}=\sqrt{\hat{s}}$.
\end{itemize}
These options coincide with those available in \mg~\cite{Hirschi:2015iia} for LO computations. The scale choice based on the transverse mass of the $2\to2$ system obtained from $k_\mathrm{T}$ clustering~\cite{Catani:2001cc} as implemented in \mg, will be added in a future release.

%%================================
\subsubsection{Python interface}
\label{sec:python}

One of the primary goals of the \madspace project is to expose as many of its core building blocks -- such as phase-space generation, adaptive sampling, and observable evaluation -- as reusable components, rather than providing only a monolithic event generator. In both theoretical and experimental workflows in high-energy physics, \python has become a common high-level language for orchestrating and combining different software tools for event generation, analysis, visualization, and machine-learning applications~\cite{Rodrigues:2020syo}. To enable the use of \madspace within this wider \python ecosystem, a lightweight and seamless interface to \python libraries for high-dimensional tensor processing, such as \numpy and \pytorch, is provided.

We implement this interface using the DLPack protocol~\cite{dlpack}, a standardized system for exchanging tensor data between different \python frameworks. DLPack supports data residing on a variety of devices, including NVIDIA and AMD GPUs, and enables different libraries to share the same underlying memory while coordinating their memory management. This avoids unnecessary data copies and reduces runtime overhead. While we currently only provide interfaces for \numpy and \pytorch, DLPack is supported by all major deep-learning frameworks in \python, and extending the interface to additional libraries such as \tensorflow or \jax in the future is straightforward.

%%%%%%%%%%%%%%%%%%%%%%%%%%%%%%%%%%%%%%%
\subsection{UMAMI -- A unified matrix element interface}
\label{sec:umami}

For interfacing between \madspace and matrix elements from the \cudacpp plugin~\cite{Hagebock:2025jyk}, we introduce the \textbf{U}nified \textbf{MA}trix ele\textbf{M}ent \textbf{I}nterface (\umami). The goal of this interface is to provide a flexible and uniform way of calling matrix-element code that can be used both for phase-space integration and for event generation within \madspace, as well as in standalone applications. This removes the need to distinguish between a dedicated “standalone” mode and an “event-generation” mode at the level of the matrix-element code, as traditionally done in \mg.

The interface operates on batches of events and supports execution on different hardware devices, including CPUs and GPUs, without requiring explicit host–device data transfers. This enables fully end-to-end GPU-resident event-generation workflows. Since most programming languages provide a C foreign-function interface, \umami is implemented as a small set of C-callable functions. Their signatures are fixed, such that \umami-compatible matrix elements can be loaded dynamically at runtime, without code generation and without requiring process-specific information at compile time. The concrete inputs and outputs of a matrix-element call are instead specified dynamically.

A call to an \umami matrix element may be as simple as requesting the squared matrix element as a function of the external momenta, but can also include additional inputs such as the value of $\alpha_s$, random numbers for helicity or colour selection, and can return auxiliary information such as the multi-channel weights. The \madspace library provides high-level, vectorized wrappers to call \umami from \python, using \numpy arrays or \pytorch tensors as input and output. In detail, the \umami interface consists of the following six functions:
\begin{itemize}
    \item \verb|umami_initialize|: initializes an independent instance of a matrix element;
    \item \verb|umami_get_meta|: queries metadata such as the target device and the numbers of external particles, Feynman diagrams, helicity configurations, and colour structures;
    \item \verb|umami_set_parameter| and \verb|umami_get_parameter|: set and retrieve model parameters;
    \item \verb|umami_matrix_element|: evaluates the matrix element. The types of inputs and outputs are specified by lists of integer keys. Possible inputs include momenta, $\alpha_s$, flavour indices, and random numbers for colour, helicity, and diagram selection. Possible outputs include the matrix element, a vector of multi-channel weights, and the selected colour, helicity, and diagram indices. The numerical data are passed as device-resident pointers with a defined memory layout;
    \item \verb|umami_free|: releases the resources associated with a matrix-element instance.
\end{itemize}
By providing the inputs, outputs, metadata fields, and supported devices as integer keys rather than encoding them in the function signatures, the interface can be extended without breaking binary compatibility. This design makes \umami directly applicable to new matrix element implementations~\cite{Frederix:2026ejl}, and future extensions to next-to-leading order (NLO), where additional inputs and outputs such as Born and virtual contributions, FKS sectors, subtraction terms, and related quantities are required. A detailed documentation of the \umami interface is provided is provided in Ref.~\cite{UMAMIdocs}.

%%%%%%%%%%%%%%%%%%%%%%%%%%%%%%%%%%%%%%%
\subsection{Unweighting and file output}

Efficient unweighting and scalable I/O are essential ingredients of modern parton-level event generators, in particular in high-statistics runs and on heterogeneous CPU/GPU architectures. In the following, we describe the strategy adopted in \madspace and contrast it with the traditional workflow in \mg, before discussing the intermediate and final event formats and the combination of multiple integration channels.

%%================================
\subsubsection{Event generation}

In \mg, event generation is split into many smaller jobs, where each job is a separate, single-threaded process responsible for generating a batch of unweighted events for one integration channel. 
The results of each job are written to Les Houches Event (LHE) files~\cite{Alwall:2006yp}. 
Only after all jobs have finished, the results are collected and a final combination and unweighting step is performed, in which the total cross section is determined and the events from all channels are merged and unweighted.
While this setup has some advantages -- most notably the complete independence of the individual jobs and therefore good scalability -- it also has several drawbacks. 
The splitting into many relatively small jobs leads to a very large number of small files, which can put a significant burden on the file system in a cluster environment. 
Communicating results via the file system and launching a separate process for each job also comes with a large runtime overhead. 
Moreover, since the results can only be combined at the very end, \mg typically generates more unweighted events than needed which then have to be discarded.

\madspace follows a different approach. 
Instead of splitting the event generation into completely independent processes, a main thread keeps track of the current integral estimate, the maximum event weight, and the number of generated events. 
This thread submits jobs to a thread pool that performs the actual event generation. 
Each job corresponds to one batch of events for a specific integration channel, and its results are returned to the main thread in memory. 
The main thread is then responsible for writing the events to disk.

\begin{figure}[t!]
    \centering
    \begin{tikzpicture}[font=\small]

\node (gen_main) [minimum height=0.5cm] {Main thread};
\node (gen_control1) [rectangle, rounded corners, fill=Ycolor, minimum width=4.3cm, minimum height=0.9cm, align=center, below=0.25cm of gen_main] {fill job queue,\\wait for results};
\node (gen_control2) [rectangle, rounded corners, fill=Ycolor, minimum width=4.3cm, minimum height=0.9cm, align=center, below=0.25cm of gen_control1] {update integral\\and max weight};
\node (gen_control3) [rectangle, rounded corners, fill=Ycolor, minimum width=4.3cm, minimum height=0.9cm, align=center, below=0.25cm of gen_control2] {unweight and write events\\to intermediate file};
\node (gen_control4) [rectangle, rounded corners, fill=Ycolor, minimum width=4.3cm, minimum height=0.9cm, align=center, below=0.25cm of gen_control3] {final unweighting step};
\node (gen_control5) [rectangle, rounded corners, fill=Ycolor, minimum width=4.3cm, minimum height=0.9cm, align=center, below=0.25cm of gen_control4] {complete LHE data\\write final output};
\draw [-Latex] (gen_control1.south) -- (gen_control2.north);
\draw [-Latex] (gen_control2.south) -- (gen_control3.north);
\draw [-Latex] (gen_control3.south) -- (gen_control4.north);
\draw [-Latex] (gen_control4.south) -- (gen_control5.north);
\draw [-Latex, thick] (gen_control3.west) -- ([xshift=-0.35cm]gen_control3.west) -- ([xshift=-0.35cm]gen_control1.west) -- (gen_control1.west);
\draw [-Latex, thick] (gen_control4.west) -- ([xshift=-0.35cm]gen_control4.west) -- node[midway, sloped, above]{if more events needed} ([xshift=-0.35cm]gen_control1.west) -- (gen_control1.west);

\node (gen_box1) [
  draw,
  rounded corners,
  line width=1pt,
  fill=none,
  inner xsep=25pt,
  inner ysep=5pt,
  fit=(gen_main) (gen_control1) (gen_control2) (gen_control3) (gen_control4) (gen_control5)] {};

\node (gen_pool) [minimum height=0.5cm, right=3.0cm of gen_main, minimum width=6cm] {CPU thread pool or GPU};
\node (gen_job11) [rectangle, rounded corners, fill=Gcolor, minimum width=3cm, minimum height=2cm, align=center, below=0.2cm of gen_pool, xshift=-1.6cm] {Generate events};
\node (gen_job12) [rectangle, rounded corners, fill=Bcolor, minimum width=3cm, minimum height=1cm, align=center, below=0.2cm of gen_job11] {Unweight\\(+ copy to CPU)};
\node (gen_job21) [rectangle, rounded corners, fill=Gcolor, minimum width=3cm, minimum height=2cm, align=center, below=0.2cm of gen_pool, xshift=1.6cm] {Generate events};
\node (gen_job22) [rectangle, rounded corners, fill=Bcolor, minimum width=3cm, minimum height=1cm, align=center, below=0.2cm of gen_job21] {Unweight\\(+ copy to CPU)};
\node (gen_box2) [
  draw,
  rounded corners,
  line width=1pt,
  fill=none,
  inner sep=5pt,
  fit=(gen_pool) (gen_job11) (gen_job12) (gen_job21) (gen_job22)] {};

\node (gen_pool2) [minimum height=0.5cm, right=3.0cm of gen_main, yshift=-4.8cm, minimum width=6cm] {CPU thread pool};
\node (gen_job13) [rectangle, rounded corners, fill=Rcolor, minimum width=3cm, minimum height=1cm, align=center, below=0.2cm of gen_pool2, xshift=-1.6cm] {generate LHE\\file content};
\node (gen_job23) [rectangle, rounded corners, fill=Rcolor, minimum width=3cm, minimum height=1cm, align=center, below=0.2cm of gen_pool2, xshift=1.6cm] {generate LHE\\file content};
\node (gen_box3) [
  draw,
  rounded corners,
  line width=1pt,
  fill=none,
  inner sep=5pt,
  fit=(gen_pool2) (gen_job13) (gen_job23)] {};

\draw [-Latex, thick] ([yshift=-0.15cm]gen_control1.east) -- ([yshift=-0.15cm]gen_control1.east -| gen_box2.west);
\draw [-Latex, thick] ([yshift=0.15cm]gen_control1.east -| gen_box2.west) -- ([yshift=0.15cm]gen_control1.east);

\draw [-Latex, thick] ([yshift=-0.15cm]gen_control3.east) -- ([yshift=-0.15cm]gen_control3.east -| gen_box2.west);
\draw [-Latex, thick] ([yshift=0.15cm]gen_control3.east -| gen_box2.west) -- ([yshift=0.15cm]gen_control3.east);

\draw [-Latex, thick] ([yshift=-0.15cm]gen_control5.east) -- ([yshift=-0.15cm]gen_control5.east -| gen_box3.west);
\draw [-Latex, thick] ([yshift=0.15cm]gen_control5.east -| gen_box3.west) -- ([yshift=0.15cm]gen_control5.east);

\end{tikzpicture}%
    \vspace{-2em}
    \caption{Illustration of the full event-generation workflow.}
    \label{fig:event_gen}
\end{figure}
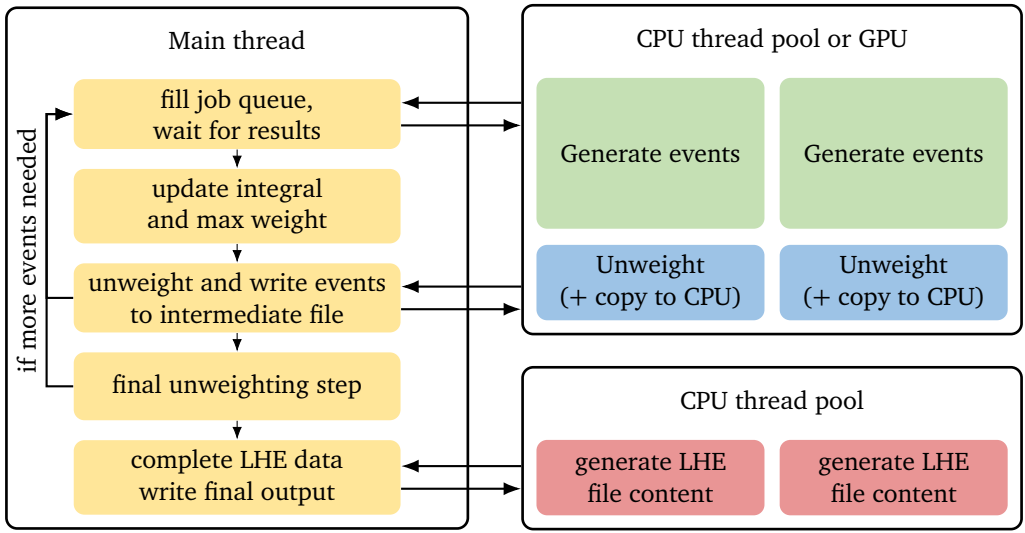

In contrast to \mg, where a separate intermediate LHE file is written for each batch of events, \madspace produces only two files per integration channel. One file stores the event weights, while the second contains the momenta and integer indices encoding the selected flavour, colour, helicity, and Feynman diagram. 
The diagram information is required, for instance, to identify resonant intermediate propagators and pass this information to the subsequent parton-shower stage, where radiation associated with the decay products of an on-shell resonance must be treated separately from additional jets produced elsewhere in the hard process. The corresponding resonance structure is therefore encoded in the LHE file and used by parton-shower programs such as \pythia or \herwig.
The channel-wise output files are unweighted using the current estimate of the per-channel maximum event weight. 
We estimate the maximum event weight on the fly, allowing for a user-defined amount of over-weight events relative to the total cross section.
The maximum weight therefore increases during the generation process, hence the events in the intermediate files are only partially unweighted. 
Once the estimated number of unweighted events reaches the target event count, a final unweighting with respect to the final maximum weight is performed. 
As the event weights are stored in a separate file, this step is fast and requires only sequential file access, avoiding costly strided I/O. 
After this step, the estimated number of unweighted events is replaced by the actual one. If it still exceeds the target event count, generation for that channel is completed. Otherwise, additional jobs are submitted to the thread pool. We illustrate the full event generation workflow in Fig.~\ref{fig:event_gen}.

%------------------------------------------------------
%------------------------------------------------------

%%================================
\subsubsection{Binary event format}
\label{sec:binary_output}

While the LHE format is useful for compatibility and for interfacing with other HEP tools, it is text-based and therefore comparatively slow to generate and parse. 
This can become a bottleneck both during event generation and in the final unweighting and combination step. 
We therefore employ a binary intermediate format. 
In order to remain interoperable with the wider \python data-processing ecosystem, the binary files are based on the \texttt{npy} format used by \numpy for storing and loading array data.

The \texttt{npy} format combines high flexibility for array-like data structures, including support for inhomogeneous data types, with a very simple underlying structure. 
In contrast to more complex formats such as HDF5, reading and writing \texttt{npy} files can be implemented without relying on external libraries, thereby reducing the number of dependencies. 
Each file consists of a header describing the data layout, including data types, column names, and array shapes, followed by the data stored as a contiguous memory block, resulting in minimal overhead for I/O operations.

We use the \texttt{npy} format both as an intermediate format and as an alternative format for the final event output. 
As described above, the intermediate files contain either the event weights as 64-bit floating-point numbers, or the momenta and associated indices as 64-bit floating-point numbers and 32-bit integers, respectively. 
The \texttt{npy}-based format for the final output contains all per-event and per-particle information specified in the LHE standard~\cite{Boos:2001cv,Alwall:2006yp}. The structure and content of these files can be inspected directly by loading them with \numpy. 
Overall, the binary representation leads to significantly faster load times and facilitates direct access to generated parton-level events with \python.

%%================================
\subsubsection{Combining channels and final event output}

After event generation is complete, the results from the individual channels have to be combined into a single output file. 
In addition, information on flavour, helicity, colour, and intermediate resonances must be reconstructed from the indices stored in the intermediate files. 
The combination is achieved by randomly selecting events from the different channels, with probabilities proportional to their relative contributions to the total cross section. 
We use look-up tables to expand the corresponding indices into the full per-particle information in an efficient manner. 
The selected events are then written to the final output file.
We support three different output formats:
\begin{enumerate}
    \item the standard XML-based LHE format;
    \item an \texttt{npy}-based binary file containing the full information of the LHE record;
    \item a more compact \texttt{npy}-based binary file containing only the momenta and indices, analogous to the intermediate format.
\end{enumerate}
This ensures compatibility to existing tools, while at the same time facilitating new \python-based workflows. 

\clearpage
%%%%%%%%%%%%%%%%%%%%%%%%%%%%%%%%%%%%%%%%%%%%%%%%%%%
\section{Validation and performance}
\label{sec:performance}

The reliability of a modern event generator cannot be judged by performance alone. It requires mathematical consistency of the underlying phase-space construction, faithful reproduction of physical distributions, and scalability in realistic production environments. The design of \madspace enables these aspects to be tested in a clean and hierarchical manner, beginning with exact checks of phase-space volumes and inverse mappings, and progressing towards full event generation and throughput benchmarks.

%%%%%%%%%%%%%%%%%%%%%%%%%%%%%%%%%%%%%%%
\subsection{Phase-space volume and inverse mappings}

%------------------------------------------
\begin{figure}[b!]
    \centering
    \includegraphics[width=0.30\textwidth]{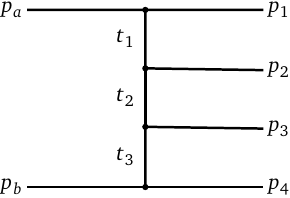}
    \hspace{3mm}
    \includegraphics[width=0.30\textwidth]{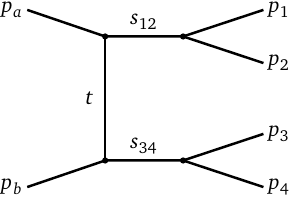}
    \hspace{3mm}
    \includegraphics[width=0.30\textwidth]{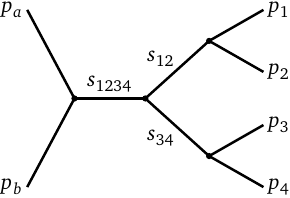}
    \caption{From left to right: A pure $t$-channel, mixed $t$- and $s$-channel, and pure $s$-channel topology for a generic $2\to4$ process. }
    \label{fig:topology_ps_volume}
\end{figure}
%------------------------------------------

\begin{figure}[t!]
    \centering
    \includegraphics[width=0.49\linewidth]{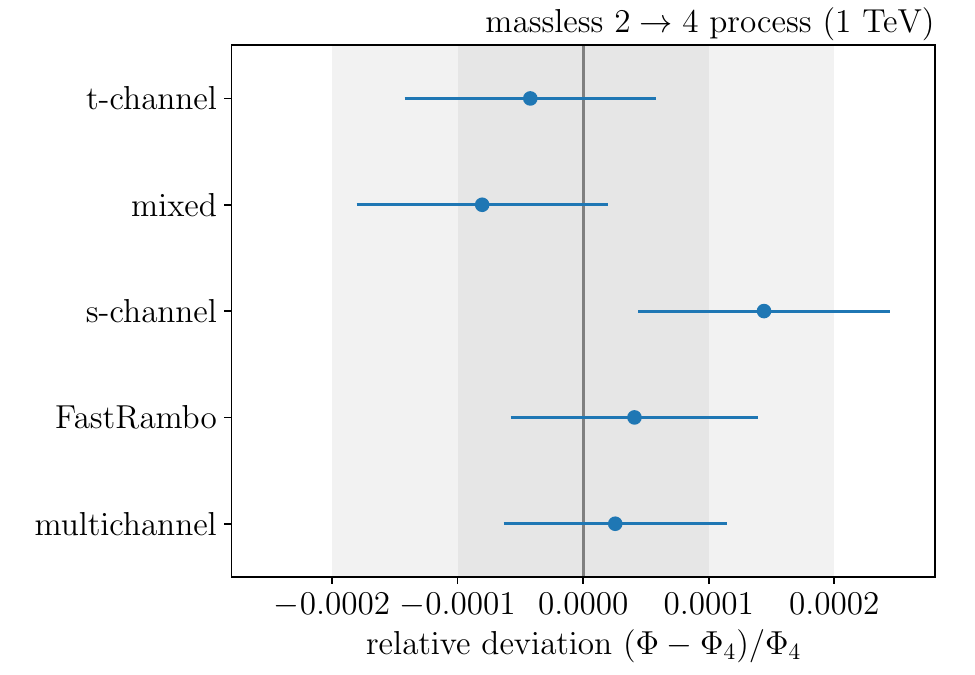}
    \includegraphics[width=0.49\linewidth]{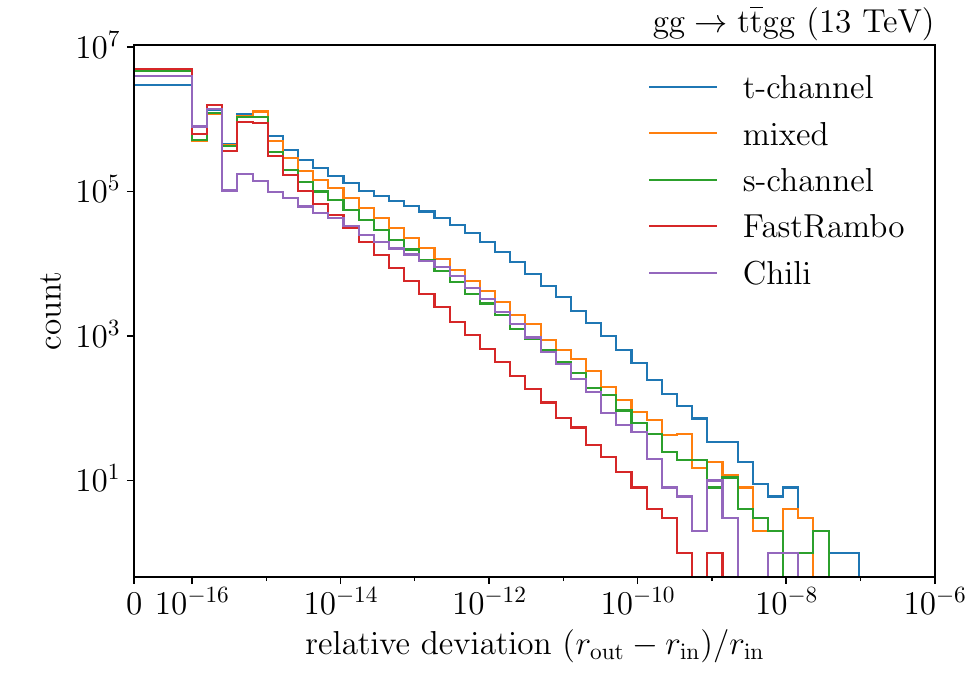}
    \caption{Left: computed phase-space volume for different mappings relative to the analytical result for a massless $2\to4$ process. Points were sampled until the target precision of $10^{-4}$ was reached. Right: histogram of the relative deviation between the random inputs and recovered random outputs when the forward and inverse mappings are evaluated subsequently for 1M points flattened over the 10 random dimensions.}
    \label{fig:ps_vol_inverse}
\end{figure}

To validate the phase-space construction implemented in \madspace, we begin by considering the phase-space volume obtained from different mappings at fixed center-of-mass energy $E_\text{CM}$, isolating the phase-space integration from matrix elements, PDFs, and flux factors. For $n$ massless final-state particles and in the absence of cuts, the analytic phase-space volume is given by
\begin{equation}
    \Phi_n(E_\text{CM},m_i=0) =  \frac{E_\text{CM}^{2n-4}}{\Gamma(n)\,\Gamma(n-1) \pi^{2n-3} 2^{4n-5}}\eqperiod
\end{equation}
We compute the phase-space volumes for a massless $2\to4$ process at $E_\text{CM} = 1\,\text{TeV}$ using five different phase-space mappings,
\begin{itemize}
    \item a pure $t$-channel topology (Fig.~\ref{fig:topology_ps_volume}, left), based on three iterative $2\to2$ scattering blocks described in Sec.~\ref{sec:two_to_two};
    \item a mixed $t$- and $s$-channel topology with one $2\to2$ scattering, followed by two $1\to2$ decays, see Sec.~\ref{sec:one_to_two}, of its outgoing legs (Fig.~\ref{fig:topology_ps_volume}, center);
    \item a pure $s$-channel topology with a first $1\to2$ decay, followed by two $1\to2$ decays of its outgoing legs (Fig.~\ref{fig:topology_ps_volume} right);
    \item the \fastrambo mapping, described in Sec.~\ref{sec:fastrambo};
    \item and a full multi-channel integration setup based on the tree-level Feynman diagram topologies for the process $\Pg\Pg \to \Pg\Pg\Pg\Pg$, including adaptive importance sampling using \vegas and channel weights based on the propagator structure of the Feynman diagrams, \ie SDE-strategy 2.
\end{itemize}
For all topology-based phase-space mappings, we use massless propagators and sample the invariants using the power-law mapping of Eq.\eqref{eq:prop_stable_nu} with $\nu = 0.3$. The mappings are automatically constructed from the phase-space building blocks using the algorithms described in Secs.~\ref{sec:s_order} and~\ref{sec:t_order}. Adaptive importance sampling is disabled for all mappings except for the full multi-channel setup. Samples are generated in batches of 10k events until the estimated relative Monte Carlo integration error falls below $10^{-4}$. The resulting phase-space volumes relative to the analytic value, shown in the left panel of Fig.~\ref{fig:ps_vol_inverse}, are in agreement with the exact result within uncertainties.

%------------------------------------------
%------------------------------------------

As a complementary consistency check, we verify that the inverse mappings correctly recover the random numbers used in the corresponding forward mappings. Specifically, we test the agreement between the random inputs $r_\text{in}$ to the forward mapping and the outputs $r_\text{out}$ returned by the inverse mapping. We consider the LO process $\Pg\Pg \to \Pt\Ptbar\Pg\Pg$, which includes both massive and massless final-state particles and requires a total of $3 \times 4 - 2 = 10$ random numbers. We use the same pure $t$-channel, mixed $t$- and $s$-channel, and pure $s$-channel topologies as in the phase-space volume test, as well as the \fastrambo and \chili mappings. We evaluate the forward and inverse mappings for 1M random inputs. The resulting distributions of the relative deviations between $r_\text{in}$ and $r_\text{out}$, flattened over the 10-dimensional random space, are shown in the right panel of Fig.~\ref{fig:ps_vol_inverse}. For the majority of phase-space points, the deviation is close to machine precision, $\mathcal{O}(10^{-16})$, with a tail towards larger deviations that approximately follows a power-law behavior. The largest deviations are observed for phase-space configurations with Mandelstam invariants $s$ and $t$ close to zero.

While the examples shown here are restricted to a limited set of representative topologies, analogous checks are performed for a much broader class of phase-space constructions as part of the automated continuous-integration (CI) test suite. These tests constitute mandatory acceptance criteria and must be passed for all code changes. They include validations of phase-space volumes, momentum conservation, on-shell conditions for outgoing particles, and the consistency of forward and inverse mappings.

%%%%%%%%%%%%%%%%%%%%%%%%%%%%%%%%%%%%%%%
\subsection{Generated phase-space distributions}
\label{sec:ps_dist}

%------------------------------------------
\begin{figure}[t!]
    \centering
    \includegraphics[width=0.32\linewidth]{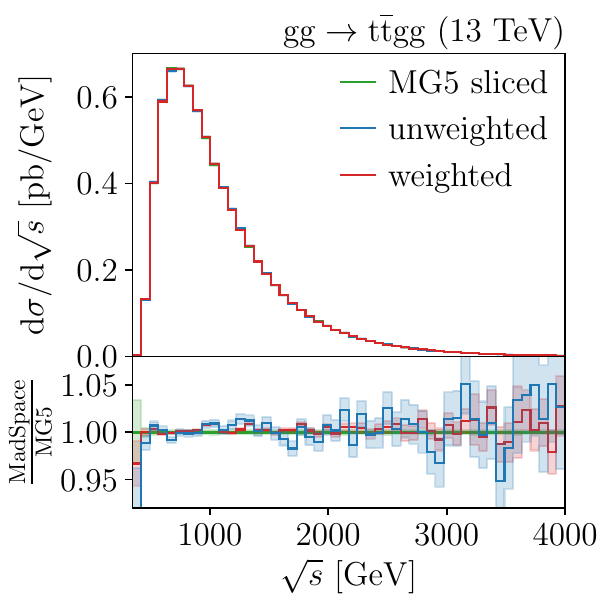}
    \includegraphics[width=0.32\linewidth]{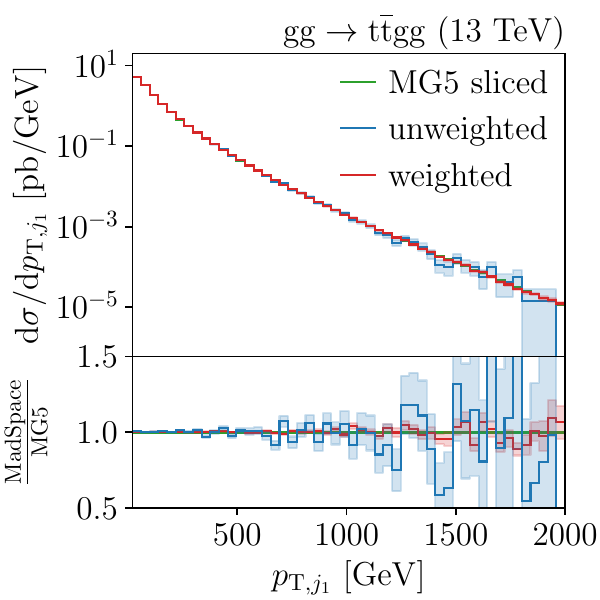}
    \includegraphics[width=0.32\linewidth]{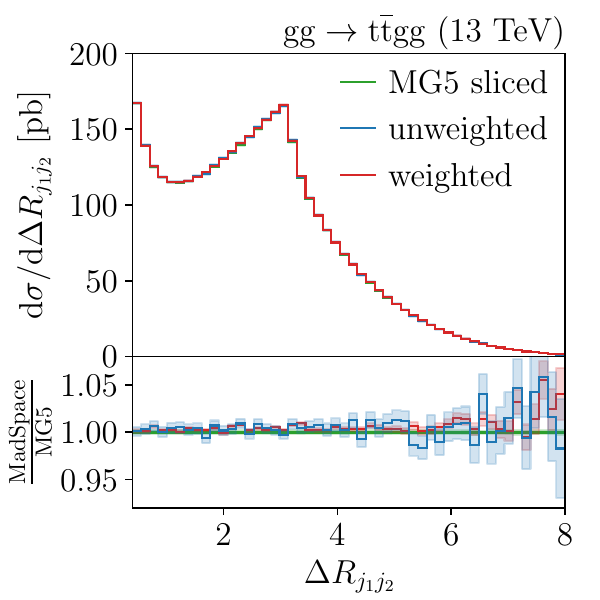}
    \includegraphics[width=0.32\linewidth]{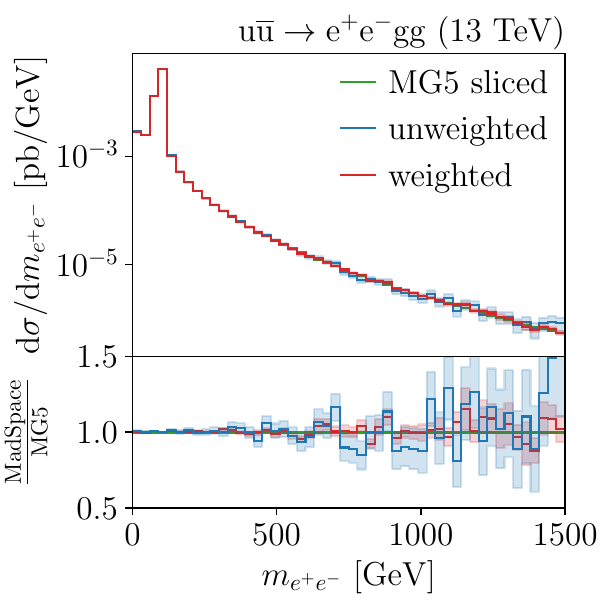}
    \includegraphics[width=0.32\linewidth]{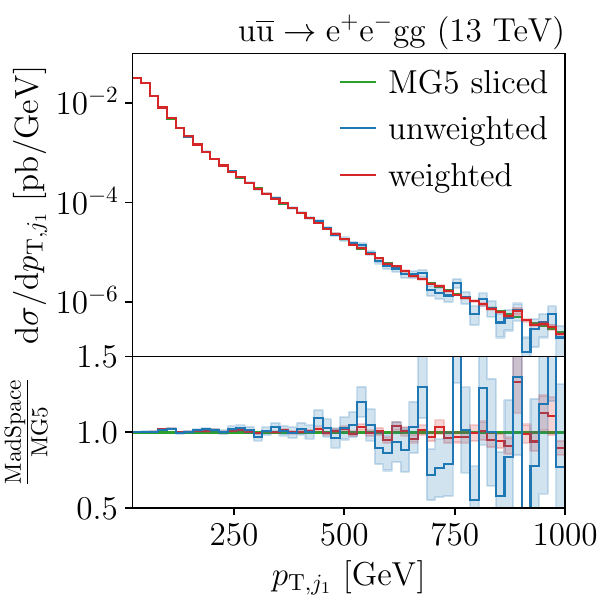}
    \includegraphics[width=0.32\linewidth]{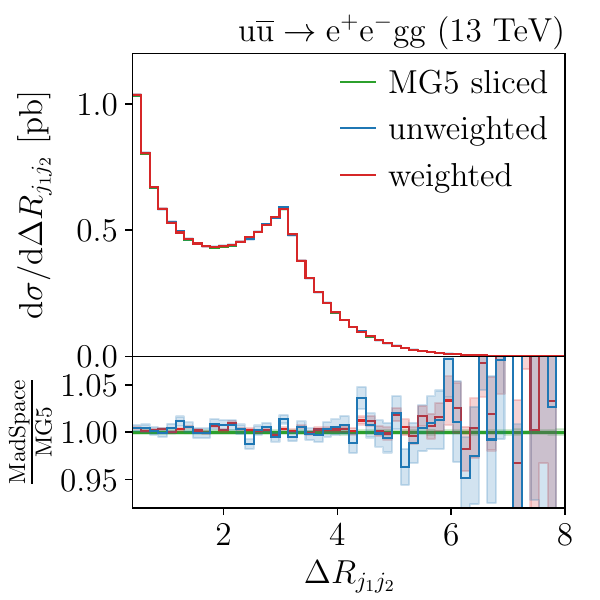}
    \caption{Differential distributions for representative phase-space observables in  $\Pg\Pg \to \Pt\Ptbar\,\Pg\Pg$ (top) and $\Pu\Pubar \to \Pep \Pem\,\Pg\Pg$ (bottom) at $\sqrt{s}=13$~TeV.  Shown are the invariant-mass (left), transverse-momentum (middle), and angular-separation (right) $\Delta R$ distributions. Results are shown for a high-statistics sliced \mg run (green), unweighted \madspace events (blue), and weighted \madspace events (red). The lower panels display the ratio of \madspace to \mg. All distributions include the MC error as error bands.}
    \label{fig:observables}
\end{figure}
%------------------------------------------

We now move to the full event-generation setup including the matrix element, PDFs, multi-channel phase space, and adaptive importance sampling with \vegas, to validate the differential distributions generated by \madspace. The \vegas grids in \madspace are optimized during an initial adaptation phase and are then frozen for event generation. Only events produced after convergence are recorded. As realistic LHC examples, we consider top-pair production $\Pg\Pg \to \Pt\Ptbar\,\Pg\Pg$ and the Drell-Yan process $\Pu\Pubar \to \Pep \Pem\,\Pg\Pg$. In both cases, we only consider the LO contribution and a single partonic sub-process with two gluon jets in the final state. 
We restrict the detailed comparisons to these two process classes because they combine massive and massless final states with resonant $s$-channel and $t$-channel structures, and because they allow for a one-to-one comparison with \mg using identical matrix elements and multi-channel setups.
Since \madspace only receives topologies, masses, and widths from \mg, other processes or UFO models do not exercise any additional functionality of the phase-space library itself.
We use a center-of-mass energy of $\sqrt{s_\text{lab}}=13$ TeV, fixed factorization and renormalization scales set to the $\PZ$ mass, $\mu_\text{F} = \mu_\text{R} = \MZ$, and the NNPDF2.3 LO PDF set~\cite{Ball:2012cx} provided by \lhapdf~\cite{Buckley:2014ana}.

For the top-pair production process, we choose the partonic center-of-mass energy $\sqrt{s}$, the transverse momentum $p_{\text{T},j_1}$ of the hardest jet, and the angular separation $\Delta R_{j_1 j_2}$ between the two jets as observables. For the Drell-Yan process, we replace the partonic COM energy with the invariant mass of the electron-positron pair. We compare three setups against each other in Fig.~\ref{fig:observables}, namely
\begin{enumerate}
    \item As a baseline, we use \mg 3.6. To ensure sufficient statistics in both the bulk and the tails of all observables, we generate 1M unweighted events each for ten equally sized slices, \ie 10M unweighted events in total, for all observables. Each slice comprises five bins in the histogram (denoted as \emph{MG5 sliced} in Fig.~\ref{fig:observables});

    \item Second, we generate 1M unweighted events total in \madspace, allowing for a $0.1\%$ over-weight events relative to the total cross section (denoted as \emph{unweighted} in Fig.~\ref{fig:observables});

    \item Finally, we also store all weighted events, about 100M events, that we sampled to obtain the 1M events in the second setup (denoted as \emph{weighted} in Fig.~\ref{fig:observables}).
\end{enumerate}
We find that both the weighted and unweighted distributions from \madspace agree well with the baseline distributions from \mg within the statistical uncertainty. While the distribution over the unweighted events has larger statistical errors due to the loss of statistics during the unweighting step, it remains unbiased even in the far tails of the distributions.

%%%%%%%%%%%%%%%%%%%%%%%%%%%%%%%%%%%%%%%
\subsection{Event-generation throughput}

%------------------------------------------
\begin{figure}[t!]
    \centering
    \includegraphics[width=0.49\linewidth]{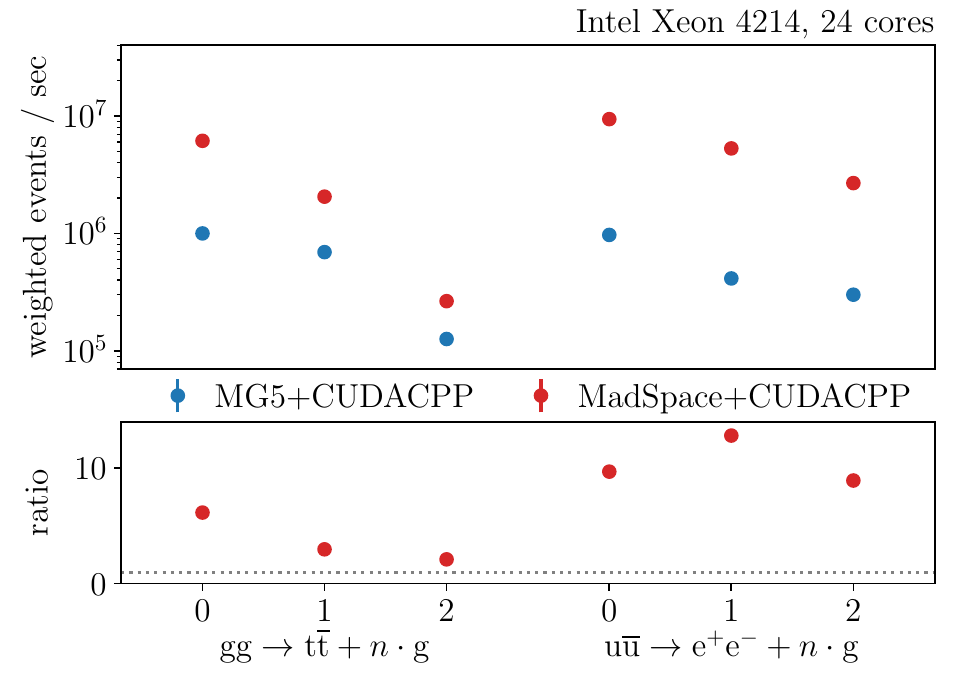}
    \includegraphics[width=0.49\linewidth]{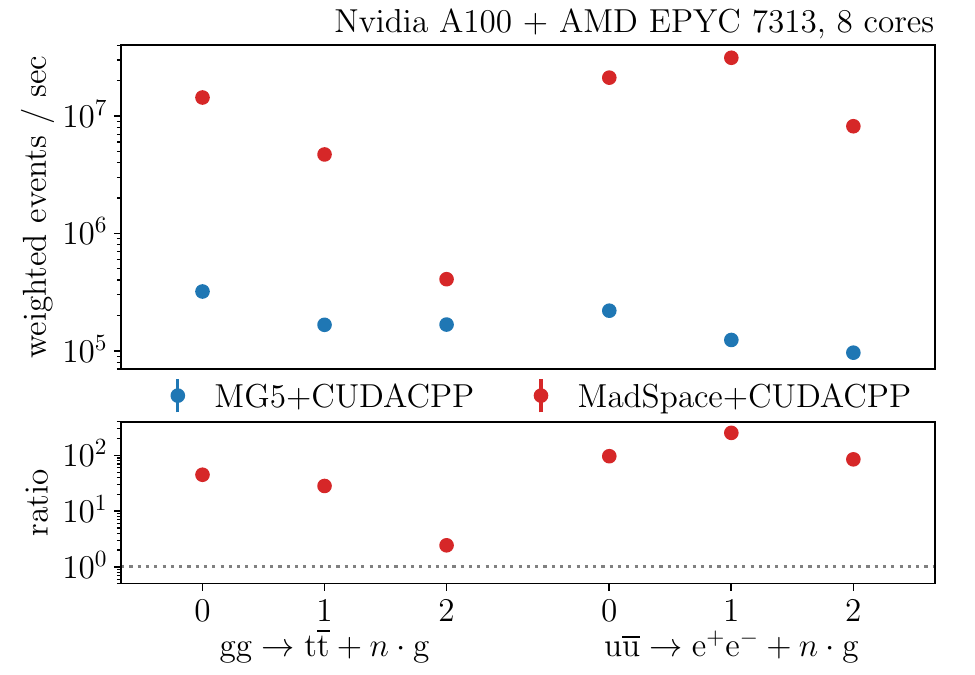}
    \caption{Throughput of \emph{weighted} event generation for representative $2\to n$ LHC processes, at LO accuracy, on CPU and GPU architectures. The upper panels show the number of weighted events per second obtained with \texttt{MG5+CUDACPPP} (blue) and with \madspace (red) for an Intel Xeon 4214 (24 cores, left) and an NVIDIA A100 GPU with AMD EPYC~7313 host (8 cores, right). Results are shown for $\Pg\Pg \to \Pt\Ptbar + n \Pg$ and $\Pu\Pubar \to \Pep \Pem + n \Pg$ for $n=0,1,2$. The lower panels display the speed-up factor. We show means and standard deviations of ten independent runs. The error bars are too small to be visible for most points.}
    \label{fig:throughput}
\end{figure}
%------------------------------------------

We benchmark the performance of \madspace in a full event-generation setup. 
As representative LHC processes, we again consider top-pair production, $\Pg\Pg \to \Pt\Ptbar + n \Pg$, and Drell–Yan, $\Pu\Pubar \to \Pep \Pem + n \Pg$, both at LO accuracy, restricting ourselves to a single partonic sub-process with up to two gluon jets in the final state.
For higher jet multiplicities, neural importance sampling with \madnis replaces \vegas as the default adaptive sampler and the relative contribution of phase-space generation and I/O becomes negligible. We therefore defer a detailed performance study in this regime to an upcoming publication on \madnis in combination with \madspace.
As a baseline, we use \mg together with the \cudacpp plugin for matrix-element evaluation. All remaining settings, including PDFs and scale choices, are identical to those used in Sec.~\ref{sec:ps_dist}. CPU benchmarks are performed on an Intel Xeon 4214 (24 cores), while GPU benchmarks use an Nvidia A100 hosted by an AMD EPYC 7313 system, restricting the number of CPU cores to eight. To ensure a fair comparison, \madspace also uses the \cudacpp matrix elements via the \umami interface, with identical multi-channel setups and \vegas as the adaptive sampler.
For the timing measurements, we only consider samples generated after convergence of the \vegas grids in \madspace, and we similarly exclude the initial survey pass in \mg. In both cases, the cost of this initial \vegas adaptation is negligible compared to the total runtime.

%------------------------------------------
\begin{figure}[t!]
    \centering
    \includegraphics[width=0.49\linewidth]{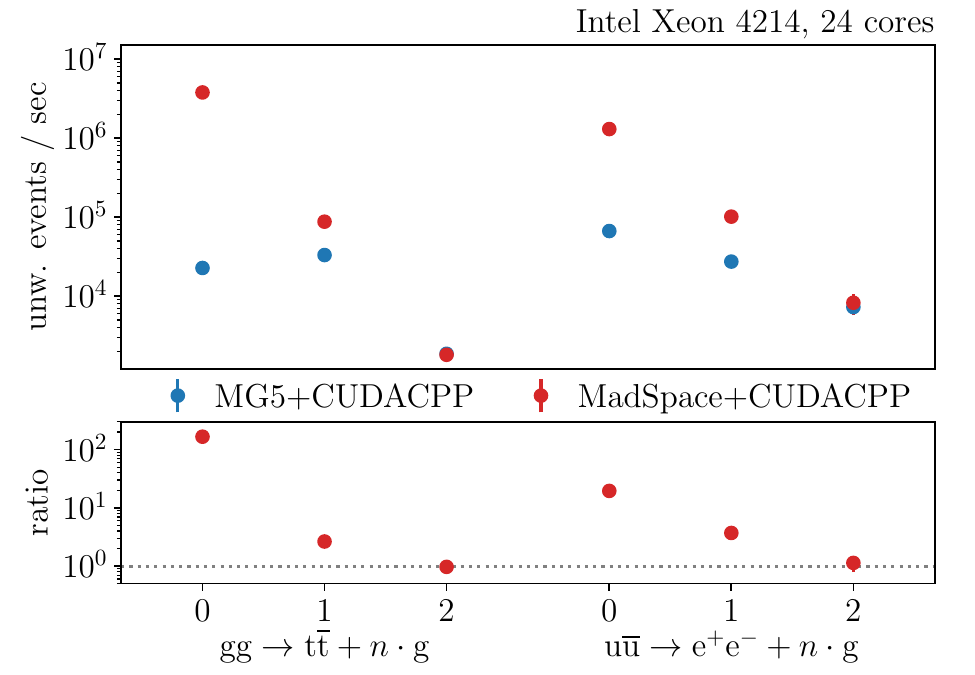}
    \includegraphics[width=0.49\linewidth]{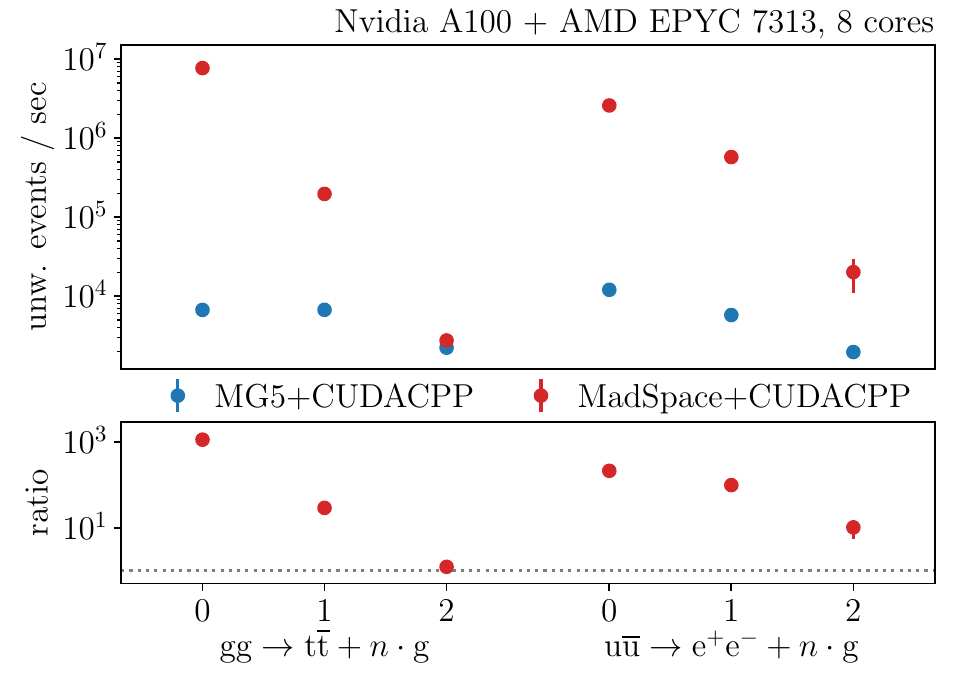}
    \caption{Throughput of \emph{unweighted} event generation for representative $2\to n$ LHC processes, at LO accuracy, on CPU and GPU architectures. The upper panels show the number of weighted events per second obtained with \texttt{MG5+CUDACPPP} (blue) and with \madspace (red) for an Intel Xeon 4214 (24 cores, left) and an NVIDIA A100 GPU with AMD EPYC~7313 host (8 cores, right). Results are shown for $\Pg\Pg \to \Pt\Ptbar + n \Pg$ and $\Pu\Pubar \to \Pep \Pem + n \Pg$ for $n=0,1,2$. The lower panels display the speed-up factor. We show means and standard deviations of ten independent runs. The error bars are too small to be visible for most points.}
    \label{fig:throughput_unw}
\end{figure}
%------------------------------------------

We first consider the weighted event-generation throughput, defined as the number of matrix-element evaluations per second, excluding phase-space points rejected by cuts. The results are shown in Fig.~\ref{fig:throughput}.
On the CPU (left panel), we observe speed-ups between 5 and 15 for processes dominated by phase-space sampling, I/O, and scheduling overheads. For matrix-element–dominated processes such as $\Pg\Pg \to \Pt\Ptbar\,\Pg\Pg$, the throughput ratio approaches unity. In absolute terms, \madspace reaches up to 10M weighted events per second, compared to about 1M for \mg.
On the GPU, the performance gap widens substantially. While \mg accelerates only the matrix-element computation, \madspace executes the entire pipeline on the GPU, from random-number generation to unweighted event production. For Drell–Yan with one additional gluon jet, we reach a throughput of roughly 30M weighted events per second, corresponding to a speed-up of about 250 relative to \mg.
As the number of available CPU cores is restricted to eight in the GPU benchmarks, the throughput of \mg is reduced compared to the CPU-only results for all processes except $\Pg\Pg \to \Pt\Ptbar\,\Pg\Pg$. This indicates that only in this case the GPU-accelerated matrix element dominates the total runtime. In contrast, \madspace uses a single CPU core only to orchestrate GPU execution and write out event data, opening the possibility for future heterogeneous event generation exploiting both CPU and GPU resources simultaneously.

%-----------------------------------
\begin{table}[b!]
\setlength{\tabcolsep}{8pt}
\centering
\begin{small}
\begin{tabular}{ll|S[table-format=3.3(1)]}
\toprule
Framework & Output format & {Time [\si{\second}]} \\
\midrule
\mg       & LHE file & 112\pm 7    \\
\madspace & LHE file & 0.76 \pm 0.02   \\
& binary format, full LHE data & 0.429 \pm 0.005  \\
& binary format, minimal       & 0.202 \pm 0.002  \\
\bottomrule
\end{tabular}
\end{small}
\caption{Time in seconds for the final combination and event output step for 1M $\Pg\Pg \to \Pt\Ptbar\,\Pg\Pg$ events on an Intel Xeon 4214 with 24 cores, excluding the time to compress the files. The timing values, and its uncertainty in brackets, were extracted from ten independent runs.}
\label{tab:combine}
\end{table}
%-----------------------------------

We next compare the unweighted event-generation throughput. In comparison to the weighted throughput, this metric also depends on the unweighting efficiency. In \madspace, we allow for a $0.1\%$ fraction of over-weight events relative to the total cross section. Moreover, the performance is affected by additional events that are generated and subsequently discarded due to inefficient job scheduling, which in \mg can amount to nearly half of the unweighted events.
The resulting unweighted throughputs on CPU and GPU are shown in Fig.~\ref{fig:throughput_unw}. In most cases, the observed speed-ups are comparable to those seen for weighted events. For $\Pg\Pg \to \Pt\Ptbar$, the unweighting efficiency in \mg is relatively low -- in fact even lower than for top-pair production with an additional jet -- leading to a large throughput ratio in favor of \madspace.
For processes with two gluon jets in the final state, \madspace exhibits a somewhat lower unweighting efficiency than \mg. This is compensated by the higher weighted-event throughput and the reduced number of discarded events. The most likely origin of this behavior is the more conservative unweighting strategy and over-weight treatment in \madspace, which yields a more reliable description of the far tails of the distributions.

Finally, we quantify the time required for the final event-output step, excluding the \texttt{gzip} compression applied by default in \mg. 
The results are shown in Tab.~\ref{tab:combine}. 
In \mg, this step is slow due to its Python implementation and the parsing of intermediate results stored in the text-based LHE format. 
In contrast, the C++ implementation in \madspace, together with the efficient binary representation of the intermediate files, leads to a speed-up exceeding two orders of magnitude. 
The combination step can be further accelerated by selecting the binary \numpy-based output format discussed in Sec.~\ref{sec:binary_output}.

\clearpage
%%%%%%%%%%%%%%%%%%%%%%%%%%%%%%%%%%%%%%%%%%%%%%%%%%%
\section{Conclusion and outlook}
\label{sec:outlook}

In this paper, we introduced \madspace, a new phase-space and event-generation library designed to address two pressing needs of modern collider simulation: scalability and performance. Its architecture is built to fully exploit massively parallel devices while retaining the flexibility and modularity of diagram-inspired multi-channel constructions.
\madspace provides a unified framework in which analytic phase-space mappings, adaptive sampling, PDF convolutions, cuts, and unweighting are expressed as a compute graph and executed efficiently on batches of events on CPUs and GPUs.

On the physics side, \madspace implements a broad set of invertible phase-space mappings.
We reviewed the multi-channel strategy with local channel weights and extended the recursive decomposition of the phase space into decay and scattering building blocks by including a double-invariant parametrization of the two-particle phase space and a genuine $1\to3$ decay block.
A second key ingredient is the availability of a fast, invertible \rambo-like mapping. Building on \rambodiet, we introduced \fastrambo which replaces the numerically expensive polynomial inversion step by an analytic rational-quadratic transformation. While this relaxes exact flatness, it increases numerical stability and preserves invertibility. This trade-off is particularly well aligned with modern workflows in which perfect flatness is rarely the limiting factor, whereas stable inverse mappings and high throughput on vectorized hardware are essential.

On the software side, the compute-graph execution model provides a compact and extensible representation of the full parton-level generation chain. Physics-specific work -- such as the analysis of diagram topologies and the assembly of channel mappings -- is performed once at graph construction time, after which the resulting byte-code graph can be executed repeatedly with negligible overhead.
We implement the graph execution on CPUs, as well as GPUs via CUDA and HIP kernels, enabling end-to-end on-device workflows in which random-number generation, mappings, PDF evaluation, cuts, weighted histograms and matrix-element kernels operate without unnecessary host--device transfers.
Instead of the split-job event-generation approach in \mg, we implement a multi-threaded setup with in-memory communication and compact binary intermediate files to reduce overhead and file-system load, while avoiding to generate substantially more events than required.
Moreover, \madspace aims to be an enabling component rather than a monolithic application. To this end, we provide a \python interface to allow seamless interoperability with tensor libraries such as \numpy and \pytorch.

To connect phase-space generation to fast matrix-element implementations, we introduced the \umami interface.
By providing a small set of fixed C-callable entry points with dynamically specified inputs and outputs, \umami enables device-resident, batched matrix-element evaluation without process-specific code generation at integration time.
This design is intended to serve both immediate use cases -- LO integration and event generation -- and future extensions that require additional inputs, outputs and metadata.

Our validation and performance studies demonstrate that these design choices translate into a practical advantage for representative LHC processes.
We verified the correctness of phase-space volumes and the numerical stability of inverse mappings, and we showed agreement of differential distributions between \madspace and high-statistics \mg reference runs for both weighted and unweighted event generation.
In throughput benchmarks on CPU and GPU systems, \madspace achieves substantial speedups in regimes where traditional generators are dominated by phase-space overhead, while the advantage naturally decreases as matrix-element costs begin to dominate for more complex final states.

\subsubsection*{Future extensions and broader impact}

The present work establishes the core framework, but several natural extensions are already within reach:
\begin{itemize}
    \item While we provide native support for \vegas and for normalizing flows within the compute graph, a systematic study of neural importance sampling in \madspace\ -- including training strategies, multi-channel mixtures, and scaling to higher-multiplicity final states -- is left to a dedicated follow-up. We expect learned samplers to become the default option for $2\to n$ processes with larger $n$, where the integrand structure cannot be captured efficiently by analytic mappings alone.

    \item A second major direction is differentiability. With invertible mappings, a compute-graph execution model, and ML-ready interfaces already in place, \madspace provides the essential ingredients for differentiable event generation and simulation-based inference. This will allow gradients of weighted observables or likelihood surrogates with respect to theory parameters (masses, couplings, PDF parameters, scale choices) to be computed efficiently, enabling gradient-based parameter estimation, profiling, and uncertainty propagation directly at the level of the generator.

    \item Due to the modular structure of \madspace, it is simple to add support for more phase-space mappings or extend the existing ones. Possible extensions include the mappings from \texttt{MadWeight}~\cite{Artoisenet:2010cn,Brochet:2018pqf} that flatten the narrow transfer function peaks appearing in phase-space integrals for the matrix element method, and improved support for applications beyond colliders, for instance through arbitrary initial-state momenta.

    \item Extending the scope beyond the current LO-focused setup will be essential. The \umami interface is designed with future NLO ingredients in mind, and the modular graph structure allows the addition of FKS-related sector decompositions, subtraction terms, and related bookkeeping while keeping all computations on the same device. 
\end{itemize}
These examples are not intended to define a closed roadmap. Instead, they illustrate directions that naturally build on the present architecture. We expect additional applications and extensions to arise from real-world usage, and we actively welcome suggestions and contributions from the community.
More broadly, \madspace is intended to be a building block for the emerging ecosystem of hardware-accelerated and ML-enhanced simulation tools. As a core component of the next major MadGraph release, it provides a path towards end-to-end hardware-accelerated event generation in which phase space, PDFs, matrix elements, and modern sampling techniques operate coherently on heterogeneous architectures.

%%%%%%%%%%%%%%%%%%%%%%%%%%%%%%%%%%%%%%%%%%%%
\section*{Acknowledgments}

We thank the full \mg team for their support, helpful comments on the paper, and valuable input throughout this work.
In particular, we are grateful to Stefan Roiser and Daniele Massaro for their efforts in integrating the \cudacpp plugin into the \madspace and \mg workflow.
OM would like to express special thanks to Fabio Maltoni and Tim Stelzer for their unwavering support over the years and for teaching him all the intricacies of the MadEvent phase‑space generator.
TH is supported by the PDR-Weave grant FNRS-DFG numéro T019324F (40020485).
OM and TH are supported by FRS-FNRS (Belgian National Scientific Research Fund) IISN
projects 4.4503.16 (MaxLHC).
Computational resources have been provided by the Consortium des Équipements de Calcul Intensif (CÉCI), funded by the Fonds de la Recherche Scientifique de Belgique (F.R.S.-FNRS) under Grant No. 2.5020.11 and by the Walloon Region.

%%%%%%%%%%%%%%%%%%%%%%%%%%%%%%%%%%%%%%%%%%%%
\appendix
\section{Installation and usage}

In the following, we show how to install \madspace and use it through its Python API. For a larger range of examples and a detailed API reference covering the Python and C++ API, we refer to the \texttt{MadGraph7} documentation website~\cite{MG7docs}.

\subsection*{Installation}

\paragraph{From PyPI}

Pre-compiled packages are available for Linux (x86\_64) and macOS (Apple silicon) for \python 3.11 to 3.14. The Linux wheels are built with the CUDA and HIP backends included.
\begin{lstlisting}[language={}, backgroundcolor=\color{codebg}]
pip install madspace
\end{lstlisting}

\paragraph{Building from source}

\madspace is developed as part of \texttt{MadGraph7}\footnote{\url{https://github.com/MadGraphTeam/MadGraph7}} and is built with the \texttt{install.py} script in its \texttt{madspace} directory. The script installs the build dependencies, locates a suitable \texttt{cmake} (3.15 or newer) and runs the build:
\begin{lstlisting}[language={}, backgroundcolor=\color{codebg}]
git clone git@github.com:MadGraphTeam/MadGraph7.git
cd MadGraph7
python madspace/install.py --source --system
\end{lstlisting}
With \texttt{{-}{-}system}, \madspace is installed into the active \python environment and can be imported like any other package. Without arguments the script is interactive and asks which backends to enable. The same choices are available as command line flags for non-interactive use, for example
\begin{lstlisting}[language={}, backgroundcolor=\color{codebg}]
python madspace/install.py --source --system --cuda --cuda-arch "75;80;86"
python madspace/install.py --source --system --hip --simd --debug
\end{lstlisting}
Run \texttt{python madspace/install.py {-}{-}help} for the full list of options.

\subsection*{Usage}

In this example, we show how to build a basic, non-adaptive integrator for $\Pu\Pdbar \to \Pmup\Pgngm\, \Pg$. This includes a phase-space mapping (Sec.~\ref{sec:ps_mappings}) for a single Feynman diagram with a resonant $\PWp$ propagator, a matrix element (Sec.~\ref{sec:umami}) and a PDF (Sec.~\ref{sec:pdf}) from the \texttt{NNPDF40\_lo\_as\_01180} set. We use a $H_\mathrm{T}/2$ scale (Sec.~\ref{sec:scale}). All steps are called from the \python interface (Sec.~\ref{sec:python}).

\paragraph{Assembling the mapping}

We map out one Feynman diagram using the recursive phase-space decomposition of Sec.~\ref{sec:fd_mappings}, with a $t$-channel quark propagator and an $s$-channel $\PWp$. This requires the $t$-channel and decay building blocks are given in Secs.~\ref{sec:two_to_two} and~\ref{sec:one_to_two}. A vertex is a triple of line labels: \texttt{i} for an incoming line, \texttt{o} for an outgoing one and \texttt{p} for a propagator, each followed by its 0-based index. The \texttt{Topology} class analyzes the phase-space topology of the diagram. The resulting mapping samples the $\PW$ invariant mass around its pole with the given mass and width, Eq.~\eqref{eq:prop_massive}, while the momentum fractions of the incoming partons are sampled as in Eq.~\eqref{eq:pdf_sampling}. Cuts (Sec.~\ref{sec:cuts}) keep the matrix element from being evaluated in the soft and collinear regions where it diverges.
\\\noindent\begin{minipage}{\linewidth}
\begin{lstlisting}[language={}, backgroundcolor=\color{codebg}]
import os

import numpy as np
import madspace as ms

E_CM = 13000.0
M_W, W_W = 80.379, 2.085
pids = [2, -1, -13, 14, 21]  # u d~ > mu+ vm g

diagram = ms.Diagram(
    [0.0, 0.0],        # incoming u, d~
    [0.0, 0.0, 0.0],   # outgoing mu+, vm, g
    [
        ms.Propagator(mass=0.0, width=0.0, pdg_id=2),   # t-channel u
        ms.Propagator(mass=M_W, width=W_W, pdg_id=24),  # s-channel W+
    ],
    [["i0", "o2", "p0"], ["p0", "i1", "p1"], ["p1", "o0", "o1"]],
)
topology = ms.Topology(diagram)

O = ms.Observable
cuts = ms.Cuts([
    ms.CutItem(O(pids, O.obs_pt, [O.jet_pids]), min=20.0),
    ms.CutItem(O(pids, O.obs_eta_abs, [O.jet_pids]), max=5.0),
])
mapping = ms.PhaseSpaceMapping(topology, E_CM, cuts=cuts)
\end{lstlisting}
\end{minipage}
Note that this mapping describes only one of the two diagrams contributing to the process (the gluon can also be radiated off the $\Pdbar$). Since the matrix element includes both and we do not employ a multi-channel decomposition (see Sec.~\ref{sec:ps_multichannel}) in this example, the estimate is still the full cross section.

\paragraph{Loading the matrix element and the PDF}

The matrix element is generated beforehand using an UMAMI-compatible matrix-element generator such as \texttt{MadMatrix}. The PDF set is installed and located through LHAPDF and then loaded with the built-in PDF interpolator.
\\\noindent\begin{minipage}{\linewidth}
\begin{lstlisting}[language={}, backgroundcolor=\color{codebg}]
proc_dir = "PROC_udx_mupvmg"
me_path = os.path.join(proc_dir, "lib", "libmadmatrix_sm_P0_udx_mupvmg_scalar_m.so")
param_card = os.path.join(proc_dir, "Cards", "param_card.dat")

ctx = ms.default_context()
api = ctx.load_matrix_element(me_path, param_card)
MEI = ms.MatrixElement.MatrixElementInput
MEO = ms.MatrixElement.MatrixElementOutput
matrix_element = ms.MatrixElement(
    api, [MEI.momenta_in, MEI.alpha_s_in], [MEO.matrix_element_out], True
)

PDF_SET = "NNPDF40_lo_as_01180"  # MadGraph7's default PDF
import lhapdf
pdf_dir = os.path.join(lhapdf.paths()[0], PDF_SET)
pdf_grid = ms.PdfGrid(os.path.join(pdf_dir, f"{PDF_SET}_0000.dat"))
pdf_grid.initialize_globals(ctx)
alphas_grid = ms.AlphaSGrid(os.path.join(pdf_dir, f"{PDF_SET}.info"))
alphas_grid.initialize_globals(ctx)
alpha_s = ms.RunningCoupling(alphas_grid)
\end{lstlisting}
\end{minipage}

\paragraph{Scale and cross section}

\texttt{EnergyScale} computes the renormalization and factorization scales from the event momenta. \texttt{DifferentialCrossSection} combines the matrix element with the flux factor, the two PDFs and $\alpha_s$, all at the chosen scale. The list \texttt{pid\_options} gives the incoming flavour combinations it accepts, here just $\Pu\Pdbar$:
\\\noindent\begin{minipage}{\linewidth}
\begin{lstlisting}[language={}, backgroundcolor=\color{codebg}]
scale = ms.EnergyScale(
    particle_count=5, type=ms.EnergyScale.DynamicalScaleType.half_transverse_mass
)
cross_section = ms.DifferentialCrossSection(
    matrix_element=matrix_element,
    cm_energy=E_CM,
    running_coupling=alpha_s,
    energy_scale=scale,
    pid_options=[[2, -1]],
    pdf1=pdf_grid, pdf2=pdf_grid,
    input_momentum_fraction=True,
)
\end{lstlisting}
\end{minipage}

\paragraph{Integrating}

The mapping gives the momenta, the momentum fractions and the phase-space Jacobian, see Eq.~\eqref{eq:ps_integral}. The cross section gives the differential rate at that point. By computing the mean and standard deviation of their product over a batch, we compute the Monte Carlo estimate of the total cross section, and its standard error:
\\\noindent\begin{minipage}{\linewidth}
\begin{lstlisting}[language={}, backgroundcolor=\color{codebg}]
rng = np.random.default_rng(0)
n = 200000
r = rng.random((n, mapping.random_dim()))
momenta, x1, x2, det = mapping.map_forward([r])
pdf_id = np.zeros(n, dtype=np.int32)  # index into pid_options; only one here

weight = cross_section(momenta, x1, x2, pdf_id) * det
sigma, error = weight.mean(), weight.std() / np.sqrt(n)
print(f"sigma = {sigma:.2f} +- {error:.2f} pb")
\end{lstlisting}
\end{minipage}
The resulting cross-section estimate is
\begin{lstlisting}[language={}, backgroundcolor=\color{codebg}]
sigma = 255.01 +- 1.71 pb
\end{lstlisting}
Because the mapping places most of the samples on the $\PW$ pole, the relative error is much smaller than with a mapping that ignores the resonance. The remaining variance can be reduced with adaptive importance sampling (Sec.~\ref{sec:adaptive}).

\clearpage
%%%%%%%%%%%%%%%%%%%%%%%%%%%%%%%%%%%%%%%%%%%%%%%%%%
\bibliography{refs}
\end{document}